\documentclass[12pt]{article}
\usepackage[body={17cm, 23cm, centered}]{geometry}
\usepackage{epsfig,verbatim,cite}
\usepackage{lmodern}
\usepackage{braket}
\usepackage[T1]{fontenc}
\usepackage{mathtext,marvosym,textcomp}
\usepackage{mathtools}
\usepackage{slashed}
\usepackage[usenames,dvipsnames,svgnames,table]{xcolor}
\usepackage{tikz}
\usepackage[normalem]{ulem}
\usepackage{tocloft}
\usepackage{epsfig,amsmath,amsfonts,amssymb,arydshln}
\usepackage[makeroom]{cancel}
\usepackage{multirow}
\usepackage{float}
 \usepackage{relsize}

 \usepackage[debug,pageanchor=false]{hyperref}
\hypersetup{colorlinks=true,linktocpage,breaklinks,
            urlcolor=blue,
            linkcolor=red,
            citecolor=blue
            }

\numberwithin{equation}{section}

    \def\e{\epsilon}
   \def\q{\theta}
   \def\m{\mu}
\def\n{\nu}    \def\r{\rho}
    \def\f{\phi}
  \def\y{\psi}

\def\F{\Phi}

\def\fr{\frac}  \def\dt{\partial}

\def\ph{\phantom}
\def\mc{\mathcal}

\def\HH{\mathbb{H}}

\def\rmU{\mathrm{U}}
\def\rmSU{\mathrm{SU}}

\newcommand\bqa {\begin{eqnarray}}
\newcommand\eqa {\end{eqnarray}}

\newcommand{\bear}{\begin{array}}
\newcommand{\enar}{\end{array}}

\newcommand{\1}{{\mu_1}}

\newcommand{\be}{\begin{equation}}
\newcommand{\ee}{\end{equation}}
\newcommand{\bea}{\begin{eqnarray}}
\newcommand{\eea}{\end{eqnarray}}

\usepackage{tikz}
\usetikzlibrary{arrows}
\usetikzlibrary{shapes.misc}
\usepackage{tikz-cd}

\usetikzlibrary{arrows,shapes,positioning, fit}
\tikzstyle{every picture}+=[remember picture]
\tikzstyle{na} = [baseline=-.5ex]
\tikzstyle{format} = [rounded rectangle,
                      thick,
                      minimum size=1cm,
                      draw=red!00!black!80,
                      top color=white,
                      bottom color=red!00!black!00,
                      on grid]
\tikzstyle{format1} = [rectangle,
                      thick,
                      minimum size=1cm,
                      draw=red!00!black!80,
                      top color=white,
                      bottom color=red!00!black!00,
                      on grid]
\tikzstyle{format0} = [rounded rectangle,
                      thick,
                      minimum size=1.2cm,
                      draw=blue!00!black!80,
                      top color=white,
                      bottom color=blue!00!black!00,
                      text centered]
\tikzstyle{formatd} = [rounded rectangle,
                      thick,
                      dashed,
                      minimum size=1cm,
                      draw=blue!50!black!80,
                      top color=white,
                      bottom color=blue!00!black!00,
                      text centered]
\tikzstyle{format1d} = [rounded rectangle,
                      thick,
                      dashed,
                      minimum size=2cm,
                      draw=blue!50!black!80,
                      top color=white,
                      bottom color=blue!00!black!00,
                      text centered]

\tikzset{cross/.style={cross out, draw=black, minimum size=2*(#1-\pgflinewidth), inner sep=0pt, outer sep=0pt},
    cross/.default={5pt}}
\usepackage[ normalem]{ulem}
\usepackage{tocloft}

\begin{document}
\renewcommand{\contentsname}{}
\renewcommand{\refname}{\begin{center}References\end{center}}
\renewcommand{\abstractname}{\begin{center}\footnotesize{\bf Abstract}\end{center}}

\vfill
\setcounter{footnote}{0}
\begin{titlepage}
\ph{preprint}

\vfill

\begin{center}
   \baselineskip=16pt
   {\large \bf  Zoo of flows  
   in a 3d gauged supergravity
   with periodic potential}
   \vskip 2cm
 Lev Astrakhantsev$^{a,b}$
 \footnote{\tt levastr@theor.jinr.ru}$^{,}$\footnote{\tt lev.astrakhantsev@phystech.su},
    Anastasia A. Golubtsova$^{a}$\footnote{\tt golubtsova@theor.jinr.ru },
     and
    Mikhail A. Podoinitsyn$^a$\footnote{\tt mpod@theor.jinr.ru}
       \vskip .6cm
             \begin{small}
                \vskip .6cm
             \begin{small}
                          {\it
                          $^a$  Bogoliubov Laboratory of Theoretical Physics, JINR, Joliot-Curie str. 6, Dubna, 141980, Russia  \\
 $^b$ Moscow Institute of Physics and Technology, Institutskii per, 9, Dolgoprudny, 141700, Russia}

\end{small}
\end{small}
\end{center}
\vfill
\begin{center}
\textbf{Abstract}
\end{center}
\begin{quote}
In this paper we construct solutions with AdS/dS asymptotics for $D=3$ truncated gauged supergravity with a periodic scalar potential. In a holographic perspective, assuming Dirichlet boundary conditions, the solutions can be interpreted as deformations of 2d dual CFTs triggered by non-zero vacuum expectation values of irrelevant operators. In addition to the domain wall type solutions, we incorporated in the analysis a black string solution, which can also be interpreted as a deformation by VEV of an irrelevant operator. Generalizing the flows to finite temperature, we find that the corresponding geometries are singular but have horizons.
For certain flows, we provide an analytical description near the horizon region. For an exact RG flow solution, we explicitly compute the Brown-York stress-energy tensor on a cutoff surface and show that the $T\overline{T}$ operator factorizes along the holographic RG flow. We also define an effective, scale-dependent deformation parameter $\mu(\phi)$, whose running is governed by the scalar field at the cutoff.
\end{quote}

\vfill
\setcounter{footnote}{0}
\end{titlepage}

\tableofcontents

\setcounter{page}{2}

\newpage
\section{Introduction}
\label{sec:intro}

The renormalization group (RG)\cite{Gell-Mann:1954yli,Bogolyubov:1956gh,Bogolyubov:1959bfo,Symanzik:1970rt,Callan:1972uj,Wilson:1974mb}
provides a powerful framework for understanding how physical theories evolve with energy scale. In this context, RG flows represent deformations of conformal field theories (CFTs), typically induced either by adding relevant operators to the action or by turning on vacuum expectation values (VEVs) for such operators.

The holographic framework provides a geometric realization of the renormalization group. Within this duality, a boundary quantum field theory and its dynamics are fully encoded in the geometry and fields of a higher-dimensional gravitational bulk \cite{Maldacena:1997re,Witten:1998qj,Gubser:1998bc}. This correspondence establishes a  precise dictionary: deformations of the conformal field theory are  mapped  to  specific asymptotically anti-de Sitter (AdS) spacetime geometries that solve the bulk equations of motion \cite{Akhmedov:1998vf,deBoer:1999tgo,Freedman:1999gp, deHaro:2000vlm,deBoer:2000cz,Bianchi:2001kw,Bianchi:2001de,Skenderis:2002wp}. In \cite{ Papadimitriou:2003is,Papadimitriou:2004rz,Papadimitriou:2004ap,Papadimitriou:2007sj,Papadimitriou:2016yit}
a systematic  and covariant approach was developed that employs an eigenfunction expansion of the dilatation operator acting on the phase space of the bulk gravitational theory.

Deformations by relevant operators $(\Delta < d)$ correspond to introducing  fields in the bulk gravitational theory, which modify the infrared (IR) geometry and trigger a flow to a new fixed point. In contrast, by irrelevant deformations $(\Delta>d)$ growing in the ultraviolet (UV), dictate the short-distance behavior of the theory and  break conformal symmetry in the UV. A generic irrelevant deformation  will back-react strongly on the AdS geometry. The holographic dual is  related to a diverging non-normalizable mode, that means a drastic deformation of the boundary CFT, potentially leading to a non-local boundary theory or a "non-AdS" asymptotic geometry. 
For certain cases, deformations caused by irrelevant operators can be classified as exotic RG flows \cite{Kiritsis:2016kog}.

An extension of the holographic renormalization 
to the case of the presence of sources of irrelevant operators was discussed in \cite{vanRees:2011fr,vanRees:2011ir}, where it was shown that to complete the analysis, one has to introduce a cut-off surface $r=r_c$. The effect of coupling integer-dimensional scalars that source irrelevant operators on the boundary theory  was explored in \cite{Broccoli:2021icm}. The thermodynamics of observables and irrelevant deformations were studied in \cite{Davison:2018nxm}. The proposal that a finite radial cutoff $r=r_c$ in AdS$_3$ corresponds to a $T\overline{T}$ deformation of the dual CFT was established in \cite{McGough:2016lol}.

Beyond AdS/CFT, holographic RG flows have also been explored in the context of the dS/CFT duality \cite{Strominger:2001pn,Leblond:2002ns,Leblond:2002tf}, where the role of the radial coordinate is replaced by time, and the flow connects Euclidean CFTs in the asymptotic past and future.

In this work, we focus on a three-dimensional truncated gauged supergravity model \cite{Deger:1999st,Deger:2000as}, which features a real scalar field with a periodic potential inherited from the full gauged supergravity action. This potential possesses multiple extrema corresponding to AdS, de Sitter (dS), and Minkowski vacua. The holographic RG flows for this supergravity with a hyperbolic potential were studied in \cite{Arkhipova:2024iem,Golubtsova:2024dad}, and Janus/RG-flow interfaces were found in \cite{Gutperle:2024yiz}. The periodic case offers a distinct and rich structure since, besides AdS extrema, the potential includes dS and  Minkowski extrema. At the same time, below we will show that all solutions in this model are related to  deformations by irrelevant operators. In the five-dimensional case, holographic flows for the $D=5$ gauged supergravity model with a cosine superpotential corresponding to deformations  of irrelevant operators were  found in \cite{Behrndt:2000kx}.

Here we construct and study solutions from AdS and dS to Minkowski and generalize them to the case of finite temperature. From the  holographic point of view, the vacua interpolating between AdS and Minkowski with the Dirichlet boundary conditions can be associated with a deformation of a 2d dual CFT by a non-zero expectation value of an irrelevant operator, which spontaneously breaks the conformal symmetry. For the general type of scalar potentials, such flows were discussed as exotic irrelevant holographic RG flows between minima of  potentials in \cite{Kiritsis:2016kog}. 

For the de Sitter case we find two types of solutions flowing to Minkowski that  differ by asymptotics near Minkowski.
Both can be interpreted as solutions from dS to Minkowski  as irrelevant deformations by VEV of a scalar operator associated with different signs of the VEV. In this case, the dual CFTs turn out to be non-unitary. An example of the $dS_{3}/CFT_{2}$ correspondence with a non-unitary CFT was discussed in \cite{Hikida:2022ltr}.

A key aspect of our analysis is the use of a dynamical systems approach to study these RG flows \cite{Gukov:2016tnp, Kiritsis:2016kog, Arefeva:2019qen,Kiritsis:2025yke}, where flows appear as trajectories in the phase space of the scalar field and the beta-function. In this work we  focus on flows, which correspond to deformations caused by the VEVs of irrelevant operators. We further generalize these flows to finite temperature, finding that the corresponding thermal geometries are mostly singular and the only  regular ones are the BTZ and Schwarzschild-de Sitter (SdS) black holes. Our findings on the singularity of generic thermal flows are consistent with an analysis from \cite{Gursoy:2018umf}. We also discuss a black string solution \cite{Deger:1999st}, showing that it corresponds to a VEV-driven deformation where the VEV is related to the mass of the black string. We define the $T\bar{T}$ operator \cite{Smirnov:2016lqw} for the exact holographic RG flow through the Brown-York stress-energy tensor and show that the operator is factorized along the flow. We also define an effective deformation parameter $\mu$, which depends on the scalar field.


The paper is organized as follows. In Section \ref{sec:model} we briefly review  $D=3$, $\mc{N}=(2,0)$ gauged supergravity model and  discuss the scalar field behavior near AdS/dS extrema of the scalar potential. In Section \ref{sec:sol} we present exact and numerical solutions flowing from AdS/dS and discuss their holographic aspects. We also come from EOMs of the model to a 2d autonomous dynamical systems  on the plane and in the disk. In Section \ref{sec:3dflowsfneq1}, we use the Poincaré transformation to project the phase  space of the system with non-zero temperature into the unit cylinder and the ball, which helps us understand the global phase portrait and allows us to construct the flows numerically. 
We also construct analytic near-horizon solutions. In Section \ref{Sect:5} we compute the Brown‑York stress‑energy tensor on a cutoff surface for the exact holographic RG flow and show that the  $T\overline{T}$-operator factorizes. We also introduce an effective deformation parameter $\mu$, which decreases along the flow. We conclude in Section \ref{sec:disc}, where we discuss our main results and sketch further directions to explore. Some technical details are left for Appendices.

\section{Setup}
\label{sec:model}
\subsection{The supergravity model}
We consider the $\mc{N}=(2,0)$ AdS$_3$ supergravity model coupled to $n$-copies of $\mc{N}=(2,0)$ scalar multiplet constructed in \cite{Deger:1999st}. The field content of this three-dimensional supergravity is given by a vielbein $e_\m{}^a$, a doublet of gravitini $\y_\m$ and a gauge field $A_\m$. The scalar multiplet consists of a complex scalar field $\F$ and a doublet of spinorial fields $\lambda$, where R-symmetry indices are omitted. In \cite{Deger:1999st} the sigma model manifold was considered to be a coset space $G/H$ with a compact and non-compact $G$ and its maximal compact subgroup $H$. In papers \cite{Golubtsova:2022hfk,Arkhipova:2024iem,Golubtsova:2024dad,Golubtsova:2024odp} the non-compact case with the hyperbolic target space $\HH^2$ was studied. In this paper we focus on a truncation of the sigma-model to that one, which includes a single complex scalar field $\F$ (that is, $n=1$) and $G$ is compact, so the target space is a two-dimensional sphere, i.e.

\begin{equation}
\begin{aligned}
\fr{\rmSU(2)}{\rmU(1)} = S^2.
 \end{aligned}
\end{equation}

In this case the bosonic part of the Lagrangian reads \cite{Deger:1999st}
\begin{equation} 
\label{lag1}
 e^{-1}\mc{L} = \fr14 R - \fr{e^{-1}}{16 m \, a^4} \e^{\m\n\r}A_\m \dt_\n A_\r - \fr{|D_\m \F|^2}{a^2(1+|\F|^2)} - V(\F), 
\end{equation}
where $e = \det e_\m{}^a$ and $D_\mu\F= (\partial_\mu-iA_\mu)\F$. Note that the parameter $m$ in \eqref{lag1} is related to the cosmological constant as $\Lambda \equiv -4m^2$ and the parameter $a$ has a non-zero value and corresponds to the curvature of the scalar manifold (sphere) \footnote{The $a=0$ case corresponds to the flat sigma model. To take this limit, first fields $\phi$ and $A_\mu$ have to be rescaled with appropriate powers of $a$, and after the limit the potential in the action reduces to a cosmological constant \cite{Deger:2000as, Izquierdo:1994jz}.}. The potential $V(\F)$ in \eqref{lag1} is given by 
\begin{equation}
\label{pot}
V(\F) =  2 {m^2} C^2 \left(2 a^2 |S|^2 -  C^2 \right) \, ,
\end{equation}
where
\begin{equation}
 C = \fr{1 - |\F|^2}{1 +  |\F|^2}, \quad S = \fr{2 \F}{1 + |\F|^2}.
\end{equation}

To bring the kinetic term of the scalar field to the standard form, one splits the complex scalar field $\F$ into its modulus $|\F| \equiv \phi$ and phase $\q$. For the former we additionally define
\begin{equation}
C \equiv \cos \f \,\,, \,\, |S| \equiv \sin \f .
\end{equation}

After this, the Lagrangian of the supergravity theory takes the form
\begin{equation}
\label{fullmodel}
e^{-1}\mc{L} = \fr14 R - \fr{e^{-1}}{a^4} \e^{\m\n\r}A_\m \dt_\n A_\r - \fr1{4a^2} \dt_\m \f \dt^\m \f - \fr1{4a^2} |S|^2 \big(\dt_\m \q + A_\m\big)\big(\dt^\m \q + A^\m\big) -  V(\f).
\end{equation}
Note that the potential depends only on the field $\f$. So from the field equations of $\q$ and the vector field $A_\m$, it is easy to see that setting $A_\m= \q = 0$ is a consistent truncation of the theory \cite{Deger:1999st}, which finally brings us to the action of the form
\be\label{act}
S = \frac{1}{4} \int d^{3}x\sqrt{|g|}\left(R -\frac{1}{a^2}(\partial\phi)^2 -4V(\phi)\right) + \frac{1}{2}\int\limits_{\partial M} d^2 x \sqrt{|\gamma|}K,
\ee
with the potential
\begin{equation}
\label{cospot}
V(\phi)=-2m^2\cos^2\phi \Big((1 + 2 a^2) \cos^2\phi - 2 a^2\Big).
\end{equation}
In \eqref{act} $K$ is  the trace of the extrinsic
curvature  and $\gamma$ is the determinant of the induced metric $\gamma_{\alpha\beta}$, with $\alpha,\beta=1,2$.

On the period $[-\pi,\pi]$ the potential \eqref{cospot} has extrema that correspond to AdS at
\be
\phi_{AdS}=c_{1}\pi,\quad c_{1}=0,\pm 1;
\ee
de Sitter extrema are located at
\be\label{dsfp}
\phi_{dS}= \pm\arccos{\left(\sqrt{\frac{a^2}{1+2a^2}}\right)}\mp c_{2}\pi, \quad c_{2}=0,1;
\ee
and Minkowski extrema on the period are
\be
\phi_{M} = \pm \frac{\pi}{2}.
\ee

\begin{figure}[ht]
\centering
\includegraphics[height=7cm]{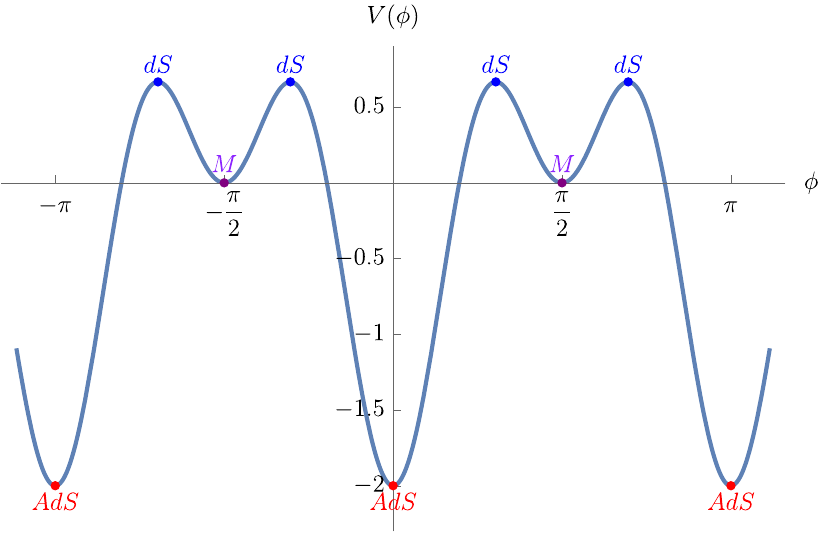}
\caption{The behavior of the potential \eqref{cospot} for $a^2=1$ and $m=1$ on the interval $[-\pi,\pi]$.}
\label{fig:cospot}
\end{figure}

We note that the number of the extrema of the potential \eqref{cospot} and their types do not depend on the parameter $a$ at all, as we can see in Fig.~\ref{fig:cospot}. Further, for our convenience, we will set $a^2=1$ when we numerically draw phase portraits and trajectories.

The potential \eqref{cospot} is related to the superpotential $W$ by the following relation:
\begin{equation}
\label{eq:V2W}
V = \fr{a^2}{4}W'^2 - \fr12 W^2,
\end{equation}
where the exact supersymmetric superpotential \cite{Deger:2002hv} has the form
\be
\label{cossuppot}
W_{\rm susy} = - 2 {m} \cos^2 \phi.
\ee

\begin{figure}[ht]
\centering
\includegraphics[height=7cm]{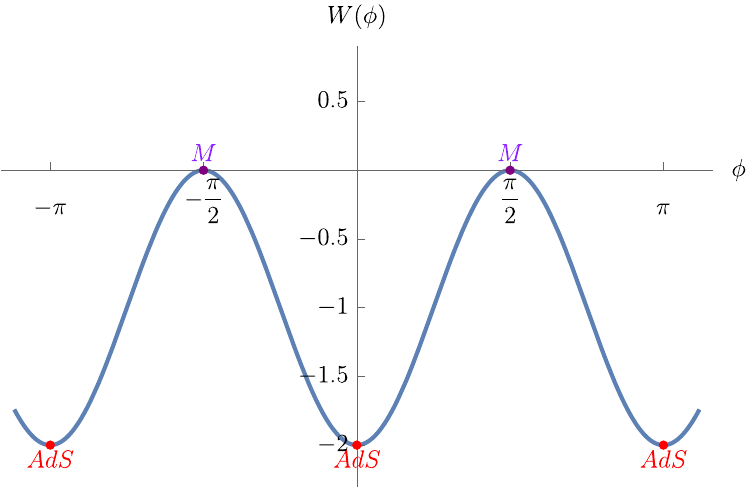}
\caption{Plot of the superpotential \eqref{cossuppot} with $m=1$. There are three $AdS$ and two Minkowski $M$ extrema on the interval $[-\pi,\pi]$.}
\label{fig:SuperPotential}
\end{figure}

The behavior of the superpotential $W$ as a function of $\phi$ is presented in Fig.~\ref{fig:SuperPotential}. Extrema of the superpotential include AdS and Minkowski vacua. 

\subsection{Equations of motion}

In this paper we will discuss solutions to \eqref{act} with  the following ansatze of the metric
\be\label{genanmet}
ds^2= e^{2A(w)} \left(-f(w)dt^2 + dx^2\right) +\frac{dw^2}{f(w)},
\ee
where $f(w)$ is the blackening function. It defines a location of the horizon, on either side of which $f$ has a different sign. A solution, which is endowed by $f$ in the ansatze, passes through a horizon and exchanges the coordinate $w$ from space-like to time-like and vice versa. 

The asymptotically BTZ black hole is given by \eqref{genanmet} with
\be
A =\sqrt{-\frac{V(\phi_{AdS})}{2}} w+\ldots
\ee
and  
\be
f = 1 -e^{-\sqrt{-2V(\phi_{AdS})}(w-w_h)}+\ldots
\ee
For $f=1$, the metric \eqref{genanmet} describes an asymptotically $AdS_{3}$ spacetime without a black hole. Extending the possible set of values for the function $f$ to the interval $[-1,0)$, we can get the asymptotically Schwarzschild-de Sitter metric  given by \eqref{genanmet} with the scale factor
\be
A =\sqrt{\frac{V(\phi_{dS})}{2}} w+\ldots
\ee
and the blackening factor
\be
f = -1 +e^{-\sqrt{2V(\phi_{dS})}(w-w_h)}+\ldots
\ee
Putting $f=-1$, we come to an asymptotically $dS$ spacetime 
\be
\label{dSfcoord}
ds^2 = e^{2A(w)}\left(dt^2+dx^2\right)-dw^2.
\ee
Note, that  using  the redefinition for the coordinates $w\to\tau$ and $t\to r$ in \eqref{dSfcoord}, we obtain the $dS_{3}$ metric in the expanding patch (for another dS coordinates representation see subsection \eqref{sec:dSsol} and appendix \eqref{appendix A}).


In what follows we will discuss  solutions to EOMs corresponding to the action \eqref{act} with the metric ansatze \eqref{genanmet}, which includes the blackening function. The Einstein equations to \eqref{act} with the metric $\eqref{genanmet}$ can be represented in the following form:
\bea
\label{eom1}
\ddot{A}+\frac{\dot{\phi}^2}{a^2}&=&0,\\
\label{eom2}
\ddot{f}+2\dot{f}\dot{A}&=&0,\\
\label{eom3}
\dot{A}\dot{f}+2\dot{A}^2f-\frac{\dot{\phi}^2}{a^2}f+4V&=&0,
\eea
where the overdot indicates the derivative with respect to the radial coordinate $w$.
 The scalar field equation is
\bea\label{eqd}
\ddot{\phi}+\dot{f}\dot{\phi}f^{-1}+2\dot{A}\dot{\phi}- 2a^2f^{-1}V_{\phi}=0,
\eea
where $V_{\phi}$ is a derivative of the potential with respect to $\phi$.

For certain flows, it will be convenient to consider the equations of motion with $f=1$. Then the equations of motion \eqref{eom1}-\eqref{eom3} take the form 
\bea\label{eqsA}
2\ddot{A}+2\dot{A}^2+4V+\frac{\dot{\phi}^2}{a^2}&=&0,\label{eom1f1}\\
2\dot{A}^2+4V-\frac{\dot{\phi}^2}{a^2}&=&0\label{eom3f1}
\eea
altogether with the scalar field equation
\bea\label{eqdf1}
\ddot{\phi}+2\dot{A}\dot{\phi}- 2a^2V_{\phi}=0.
\eea


 Note that for $f=1$, from the variation of the gravitini and dilatino, we have the following first-order  BPS equations \cite{Deger:2002hv}
\begin{equation} \label{first}
    \begin{aligned}
        \dot A &= - W, \\
        \dot \f &= a^2 W'. 
    \end{aligned}
\end{equation}
 In the derivation of eqs. \eqref{first} in \cite{Deger:2002hv}, the constraint of the preservation of supersymmetry was used. The solutions to eqs. \eqref{first} also solve the second-order equations of motion for \eqref{act} with \eqref{cospot}. However, the superpotential does not have dS extrema, solutions to \eqref{first} interpolate  between AdS and Minkowski spacetimes.

\subsection{The scalar field near AdS and dS regions}

Near AdS extrema, which correspond to $\phi_{AdS}=c_{1}\pi$, with $c_{1}=0,\pm 1,$ the scalar potential has the following expansion: 
\be
V= -2+4\big(1+a^2\big)\big(\phi-\phi_{AdS}\big)^2+\dots\,.
\ee
The AdS curvature and the mass are given, correspondingly,
\be\label{massads}
\ell^2=-\frac{1}{2V(\phi_{AdS})}=\frac{1}{4},\quad M^2
=16a^2(1+a^2).
\ee

The central charge of the dual theory is related to $\ell$ via the Brown-Henneaux formula
\be\label{ccharge}
c=\frac{3\ell}{2 G_{N}}.
\ee

Due to the form of the potential \eqref{cospot}, the AdS spacetimes for any $\phi$ such that $V'(\phi)=0$ have the same radii, and this is also relevant for the case of dS.

The generic solution to the scalar field near AdS regions has the following form:
\be\label{genscalsol}
\phi \sim \phi_{-}e^{-\Delta_{-}w}+ \phi_{+}e^{-\Delta_{+}w}\,,
\ee
where the corresponding scaling dimensions $\Delta_{\pm}$ are given by:
\be\label{deltaads}
\Delta_{\pm}
=1\pm(1+2a^2).
\ee

In the standard holographic dictionary, $\phi_{+}$, $\phi_{-}$ from  \eqref{genscalsol} are related to the VEV of the dual operator and the source, correspondingly. 
For our model that $\Delta_{-}$ \eqref{deltaads} is always negative for any $a^2>0$, while $\Delta_{+}>d$, i.e. $\Delta_{+}$ is irrelevant.
In particular, for $a^2=1$ we have $\Delta_{+}=4$ and $\Delta_{-}=-2$. 

The fact that $\Delta_{-}$ \eqref{deltaads} is far below the unitarity bound  precludes the use of Neumann and mixed boundary conditions. For the bulk scalar with  the mass $M$ given by \eqref{massads}, the Dirichlet boundary condition is the only admissible one, leading to a unitary CFT dual with a scalar operator of dimension $\Delta_{+}=2(1+a^2)$.
Since $\Delta_{+}>d$ for any $a^2>0$, the corresponding deformation is irrelevant, we can assume the dual theory is defined on the cut-off surface $w=w_c$, so Dirichlet b.c. read  
\be\label{Aphiwc}  
\phi \sim \phi_{+} e^{-\Delta_{+}w}.
\ee


Near dS extrema \eqref{dsfp}, the potential has the following form:
\be
V=\frac{2a^4}{1+2a^2}-\frac{8a^2(1+a^2)}{1+2a^2}\big(\phi-\phi_{dS}\big)^2+\dots
\ee
The parameter $H$ and the mass $M$ for the de Sitter case are defined by
\be
H^2=\frac{4V(\phi_{*})}{2}=\frac{4a^4}{1+2a^2},\quad
 M^2=2a^2V''(\phi_{*})=-\frac{32a^4(1+a^2)}{1+2a^2},
 \ee
so for any $a^2>0$  we have $M^2<0$, i.e. the scalar field is tachyonic (for tachyons in dS see \cite{Epstein:2014jaa}).
 The scaling dimensions \cite{Strominger:2001pn} are defined by
 \be
h^{\pm}=\frac{d}{2}\pm\sqrt{\frac{d^2}{4}-\frac{M^2}{H^2}}=1\pm\sqrt{9+8a^2}.
\ee

Similar to the AdS case, here, $h_{-}$ is  negative for any $a^2>0$, while $h_{+}>d$. Though the negative mass squared seems admissible in $dS$ satisfying 
 the analog of the BF bound $M^2<\frac{d^2}{4}H^2$ \cite{Kiritsis:2019wyk,Kiritsis:2025ytb}, nevertheless the scalar field is tachyonic in this case, so a holographic dual theory is non-unitary. It is interesting to note that for $a^2=2$, $\Delta_{+}$ have the same value for both AdS and dS cases and are equal to $\Delta_{+}=6$.

The solution to the scalar field near $dS$ with the metric in the form \eqref{dSfsc} reads \cite{Chernikov:1968zm,Tagirov:1972vv,Bousso:2001mw, Balasubramanian:2002zh}
\be\label{dsscalgen}
\phi \sim \phi_{dS} + \phi^{+} e^{-2h^{+}t}+ \phi^{-}e^{-2h^{-}t},
\ee
with $t\to +\infty$. Due to  $h_{-}$ is negative for all real $a^2>0$ the second term in \eqref{dsscalgen} is  apparently divergent which corresponds to unbounded energy for tachyons in non-unitary CFT. The central charge for  the dS case can be related with $H$ using \eqref{ccharge} and  $H=i\ell$ \cite{Strominger:2001pn,Collier:2025lux}. Note that in \cite{Skenderis:2002wp} it was shown that the results of the near-boundary analysis of asymptotically AdS spacetimes can be  applied to asymptotically de Sitter spacetimes.

\section{Flows from AdS/dS to Minkowski}
\label{sec:sol}
\subsection{Exact flows driven by VEV of an irrelevant operator}

In this subsection we will discuss two exact solutions to the action \eqref{act}, which solve the first-order equations \eqref{first}.

\subsubsection{Domain wall solution}
\begin{figure}[t]
    \centering
    \includegraphics[height=9cm]{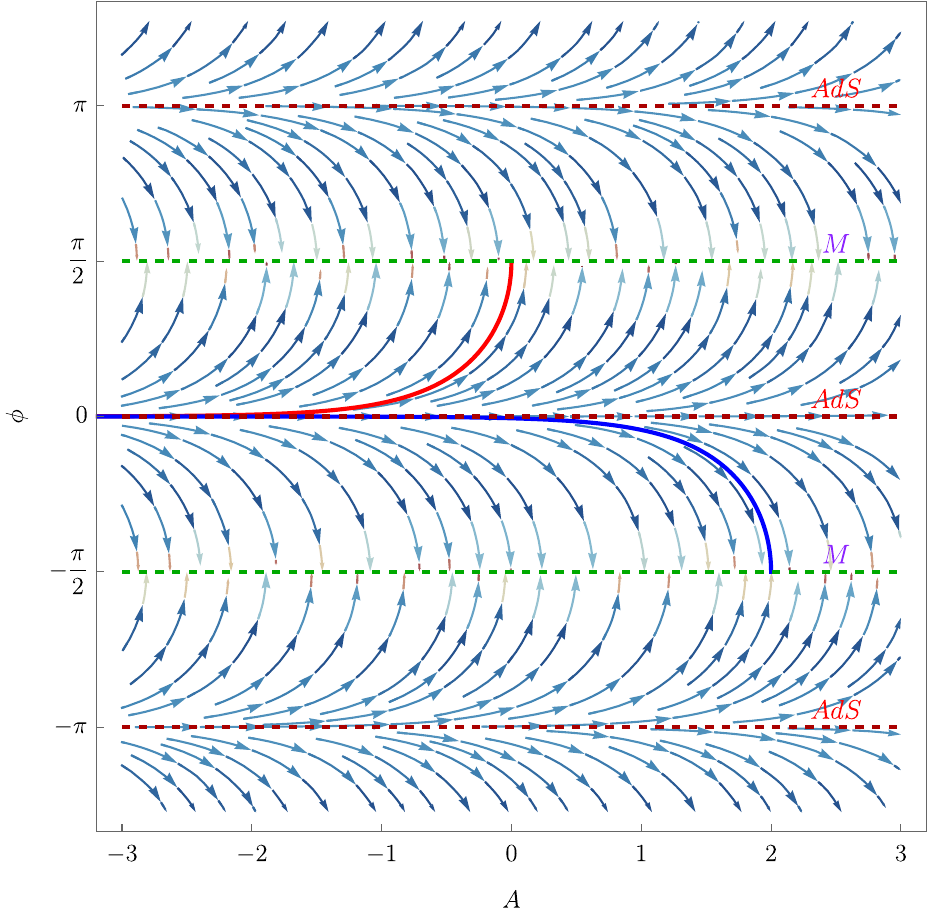}
    \caption{Phase portrait of the supersymmetric system \eqref{first}, with  $m=1$, $a^2=1$. Red and blue trajectories correspond to the same $n=0$, but to opposite signs for the dilaton field, together with different constants $\mathrm{c_A}$ for the scale factor of the exact solution \eqref{Degsol}.}
    \label{fig:FirstOrderFlow}
\end{figure}

There is an exact half-supersymmetric solution to (\ref{first}), which starts at AdS as $w\to-\infty$ and ends at Minkowski as $w\to\infty$, which both correspond to the minima of the scalar potential. The scale function $A$ and scalar field $\phi$ have the following dependence on the radial coordinate $w$:
\be \label{Degsol}
A(w) = -\frac{1}{4 a^2}  \ln \big(e^{-8 m a^2 w} +1\big) +\mathrm{c_A} \,, \qquad 
\phi = \pm \arctan \big(e^{4 m a^2 w}\big) \pm n \, \pi \,, 
\ee
where $\mathrm{c_A} \in \mathbb{R}, \, n \in \mathbb{Z}$ and $w\in (-\infty, \infty)$.  
In Fig.~\ref{fig:FirstOrderFlow} we show possible supersymmetric flows of \eqref{first} with $a^2=1$ and $m=1$.

The exact solution \eqref{Degsol} solves the  first order equations including the superpotential $W$ \eqref{cossuppot}, with  the Dirichlet boundary conditions \eqref{Aphiwc} imposed near the AdS boundary
\be
\phi \sim \phi_{+}e^{-\Delta_{+}w}+\ldots,
\ee
i.e. $\phi_{-}=0$. This asymptotic behavior can  be seen from the expansion  of the  exact solution for the scalar field \eqref{Degsol}.  Thus, the solution is dual to a deformation of CFT to a gapped phase, which  is being driven by the VEV $\phi_{+}$ of an irrelevant operator. Solutions of this type were discussed in the classification of exotic RG flows  in \cite{Kiritsis:2016kog}. Moreover,  flows triggered by non-zero VEV of irrelevant operators were discussed in  the context of deformations in the Coulomb branch in \cite{Bobev:2009ms}.

\subsubsection{Black string}
In \cite{Deger:1999st}, a black string solution was constructed for the model \eqref{lag1} with a single scalar field, which parametrizes the target  space $S^2$. The black string metric is given by
\be\label{mst}
ds^2=\left(1-\frac{M}{r}\right)^{\frac{1}{2a^2}}(-dt^2+dx^2)+\frac{1}{64m^2a^4r^2}\left(1-\frac{M}{r}\right)^{-2}{dr^2},
\ee
where $M$ is a mass  and the horizon is located at $r_h=M$.
The scalar field of the black string solution is given by
\be\label{phipphim}
\Phi_{+} =\left(\frac{\sqrt{r}-\sqrt{M}}{\sqrt{r}+\sqrt{M}}\right)^{1/2}, \quad \Phi_{-} =\left(\frac{\sqrt{r}+\sqrt{M}}{\sqrt{r}-\sqrt{M}}\right)^{1/2}.
\ee
In \eqref{phipphim}, the fields $\Phi_{+},\Phi_{-}$ are the stereographic coordinates of $S^2$ such that $0\leq\Phi_{+}\leq1$ and $1\leq\Phi_{-}\leq\infty$. The fields $\Phi_{+}$ and $\Phi_{-}$ constitute a well-defined map on the upper and lower hemispheres, $S_{+}^{2}$ and $S_{-}^{2}$, accordingly. 

Doing the scalar field transformation, we come to
\be\label{scst}
\phi=2\arctan{\Phi}=2\arctan\left(\frac{\sqrt{r}-\sqrt{M}}{\sqrt{r}+\sqrt{M}}\right)^{1/2},
\ee
where we choose the scalar field on the upper hemisphere. 

The Hawking temperature of the black string solution \eqref{mst}-\eqref{scst} can be calculated as follows:
\begin{equation}
    T_{H}
    =\frac{2ma}{\pi}\sqrt{\frac{M}{r}\Big(1-\frac{M}{r}\Big)^{\frac{1}{2a^2}}}\Bigg|_{r=M}=0.
\end{equation}
The vanishing temperature indicates that the horizon of the black string solution is degenerate, so the solution is extremal. Extremal charged black string solutions in three dimensions were studied in \cite{Kaloper:1998vw}.

The metric of the black string can be rewritten in the form \eqref{genanmet} with $f=1$. This is achieved 
by a suitable reparameterization of the radial coordinate: 
\be\label{chcbst}
\frac{dr}{8a^2 r\left(1-\frac{M}{r}\right)} =dw, \quad w= +\frac{1}{8a^2}\ln\big(r-M\big), \quad r =M+e^{8a^2w},
\ee
where we put $m^2=1$.
Then, taking into account \eqref{chcbst}, the black string metric in terms of the $w$ coordinate has the form
\be
ds^2=\left(\frac{e^{8a^2w}}{M+e^{8a^2w}}\right)^{\frac{1}{2a^2}}\big(-dt^2+dx^2\big)+dw^2,
\ee
with the scalar field given  by
\be
\label{bsphi}
\phi(w) = 2\arctan\left(\frac{\sqrt{M+e^{8a^2w}}-\sqrt{M}}{\sqrt{M+e^{8a^2w}}+\sqrt{M}}\right)^{1/2}=\pm\arctan\left(\frac{1}{\sqrt{M}}e^{4a^2w}\right),
\ee
and the scale factor reads
\be
\label{bsscaleA}
\quad A(w)=-\frac{1}{4a^2}\log\left(1+Me^{-8a^2w}\right).
\ee
One can check that \eqref{bsphi}-\eqref{bsscaleA} solves the first-order equations \eqref{first} with the supersymmetric potential \eqref{cossuppot}.

Compared \eqref{bsphi}-\eqref{bsscaleA} to the domain wall solution \eqref{Degsol}, the black string differs by the factor $M$. 
The scalar field near $w\to -\infty$ has the expansion
\be
\phi(w\to -\infty)=\pm\frac{1}{\sqrt{M}}e^{4a^2w}+\cdots,
\ee
that corresponds to the
contribution of the term with non-zero VEV $\phi_{+}$ of the irrelevant operator with the scaling dimension $\Delta_{+}=4a^2$.
Thus, the black string can be interpreted as a deformation triggered by a non-zero VEV of an irrelevant operator. 
The parameter $M$ plays the role of a modulus controlling the scale of the condensate, i.e.,
\be
\Braket{\mathcal{O}}\propto \frac{1}{\sqrt{M}}.
\ee

Near  Minkowski with $w\to +\infty$, the scalar field has an expansion
\be
\phi(w\to +\infty)=\frac{\pi}{2}\mp\sqrt{M}e^{-4a^2w}+\cdots,
\ee
which coincide with $\eqref{phiMinksol}$ for $c_{1}=0, c_{2}=\mp\sqrt{M}$ and $a^2=1$.

\subsection{Flows with $f=1$ through dynamical system}
\label{sec:2dflowsf1}

In this section, following the works \cite{Golubtsova:2022hfk,Arkhipova:2024iem}, we derive autonomous dynamical systems from the gravity equations of motion for the case $f=1$. We consider dynamical systems with phase spaces $\mathbf{R}^2$ and $\mathbf{D}^2$. The latter will clarify the dynamics near extrema points corresponding to Minkowski spacetimes. First, we construct numerical trajectories for $\phi$. Then we place the system in the disk $\mathbf{D}^2$ to properly explore the asymptotic dynamics.

\subsubsection{Dynamical system on the plane  $\mathbf{R}^2$}

To come to an autonomous dynamical system, it is convenient to use the scalar variable $X$ \cite{Gursoy:2008za}
\be
X=\frac{\dot{\phi}}{\dot{A}},
\ee
which was also used for the case of the hyperbolic target space of the sigma-model \cite{Golubtsova:2022hfk,Golubtsova:2024dad,Golubtsova:2024odp}.
Then we can represent the equations of motion as the autonomous system
\begin{equation}
\label{cossys}
    \begin{aligned}
        \frac{d\phi}{dA} & =   X , \\
        \frac{dX}{dA} & = \bigg(\frac{X^2}{a^2} - 2 \bigg)\bigg(X + \frac{a^2}{2} \fr{V'}{V} \bigg), 
    \end{aligned}
\end{equation}
where the scalar field $\phi$ is defined on the interval $[-\pi,\pi]$. 

The system \eqref{cossys} has equilibrium points with coordinates $(\phi,X) =(0,0)$ and $(\phi,X) =(\pm\arccos{\left(\sqrt{\frac{a^2}{1+2a^2}}\right)},0)$, which are correspondingly related to AdS and dS vacua. It is important to note that the Minkowski vacua with $\phi=\pm\frac{\pi}{2}$, despite being the extrema of the potential, are not critical points of the system $\eqref{cossys}$, since $X$ is divergent in this case. Note that all equilibrium points of the system \eqref{cossys} are only saddles, so the phase portrait consists  of stable/unstable manifolds. The manifolds act as separatrixes.

\begin{figure}[ht]
    \centering
    \includegraphics[height=12cm]{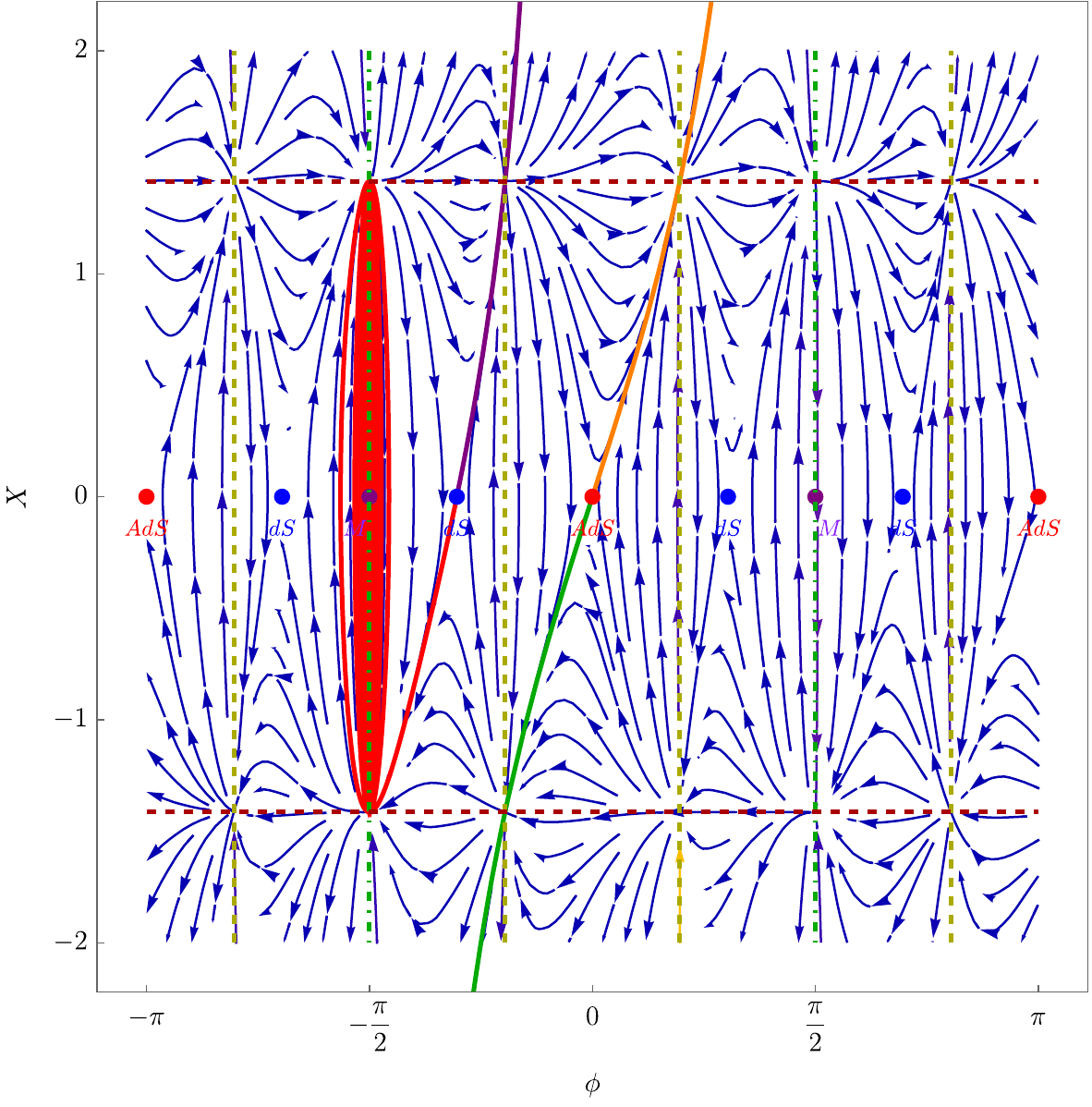}
    \caption{Phase portrait and numerical trajectories on the period $\phi\in(-\pi,\pi)$. We set $a^2=1$. Orange and green curves from the $AdS$ point are two branches of supersymmetric explicit  solution \eqref{Degsol}. Red and purple curves are the numerical solutions of the system \eqref{cossys} with initial conditions $(\phi,X) = (-0.955,\pm 0.025)$.}
    \label{fig:cosflow}
\end{figure}

The phase portrait of the system $\eqref{cossys}$ is depicted in Fig.~\ref{fig:cosflow}.
Here and in the following, by using red and blue points, we explicitly label the $AdS$ and $dS$ points correspondingly. By purple points $M$ we formally label the Minkowski space, but these points exist on the phase portrait implicitly because they coincide with the points where $V=0$, where the vector field is singular. In Fig.~\ref{fig:cosflow}, four yellow and two green dashed vertical lines correspond to the zeroes of the potential, and horizontal lines are related to the equality $X^2=2a^2$, compensating the divergence in the term $V_{\phi}/V$. Hence, at the intersections of horizontal with vertical lines, the vector field is regular. As previously in \cite{Arkhipova:2024iem}, these intersection points are similar to the phase intersection points of an ordinary pendulum.

In Fig.~\ref{fig:cosflow}, from the $AdS$ equilibrium point, we observe flows, which correspond to branches of the exact supersymmetric flow \eqref{Degsol}. We show the branches of the exact flow from AdS to Minkowski by orange and green curves in Fig.~\ref{fig:cosflow}. The flows are also presented in Fig.~\ref{fig:FirstOrderFlow} corresponding to the first-order system $\eqref{first}$. In Fig.~\ref{fig:cosflow}, the branches of the flow, starting from the $AdS$ saddle point, run to $X\to\pm \infty$, getting closer to singular lines, which are shown by red and represent Minkowski points. From the exact solution, we know that this flow should end in the Minkowski point, but from Fig.~$\ref{fig:cosflow}$ corresponding to the system \eqref{cossys}, we cannot see it precisely. 

Linearizing the dynamical system (\ref{cossys}), we can find the eigenvalues and eigenvectors and then write down an asymptotic solution. The general solution near $AdS$ at $\phi=0$ can be presented in the following form:
\bea
\label{Luapsum}
\left[\begin{array}{c}
\phi\\
X
\end{array}\right]=\left[\begin{array}{c}
0\\
0
\end{array}\right]+k_{1}e^{-\Delta_{-}A}u_{1}+k_{2}e^{-\Delta_{+}A}u_{2},
\eea
where the scaling dimensions $\Delta_{\pm}$ \eqref{deltaads} coincide with the eigenvalues $\lambda_{1,2}$ taken with opposite signs, $\Delta_{\pm}=-\lambda_{1,2}$, the eigenvalues are $\lambda_{1,2}=-1\pm (1+2a^2)$, and the eigenvectors read
\be
u_{1} =\left[\begin{array}{c}
1\\
\lambda_{1}\end{array}\right],\quad u_{2} =\left[\begin{array}{c}
1\\
\lambda_{2}\end{array}\right].
\ee

Let us remind from \cite{Arkhipova:2024iem} that setting $k_{1}$ or $k_{2}$ to zero in \eqref{Luapsum} can be associated with  imposing the Dirichlet or Neumann boundary conditions on the dilaton, correspondingly. Here, as we see, $\Delta_{-}<0$, which makes $e^{-\Delta_{-}A}$ divergent and \eqref{Luapsum} invalid for large $A$. It means that we cannot impose the Neumann b.c. in this case, i.e., we should always keep $k_{1}=0$, imposing the Dirichlet b.c. instead. Hence, on the quantum theory side, we can interpret the phase flows only as they are triggered by non-zero VEV of some irrelevant operator.

From $dS$ equilibrium points, which are saddles, we see two numerical flows in Fig.~\ref{fig:cosflow} shown by the purple and the red solid curves. The first one tends to the singular line, crosses the red dashed line, and goes to $X\to \infty$. Again, in this region, we also cannot determine its asymptotic behavior due to an unsuitable coordinate system.  We will find its asymptotic behavior considering the dynamical system in the disk $\mathbf{D}^2$ below. However, we can find an asymptotic behavior near $dS$ using Lyapunov analysis. Thus, near $dS$ we have the following general solution:

\bea\label{dSdynsys}
\left[\begin{array}{c}
\phi\\
X
\end{array}\right]=\left[\begin{array}{c}
\pm\arccos\left(\sqrt{\frac{a^2}{1+2a^2}}\right)\\
0
\end{array}\right]+k_{1}e^{-\Delta_{-}A}u_{1}+k_{2}e^{-\Delta_{+}A}u_{2},
\eea
where $\Delta_{\pm}=-\lambda_{1,2}$ with the eigenvalues given by $\lambda_{1,2}=-1\pm\sqrt{9+8 a^2}$ and the eigenvectors are 
\be
u_{1} =\left[\begin{array}{c}
1\\
\lambda_{1}\end{array}\right],\quad u_{2} =\left[\begin{array}{c}
1\\
\lambda_{2}\end{array}\right].
\ee

The second flow from $dS$ is shown by the red solid curve in Fig.~\ref{fig:cosflow}. This curve starts from $dS$ and tends to the singular line corresponding to the null of the potential coinciding with the Minkowski extremum. As we see in Fig.~\ref{fig:cosflow}, the red curve is going in converging ellipses around the Minkowski point $M$. The size of the major axis is constant, while the minor axis gradually decreases to a very small size. We can write down this solution analytically. For that, we expand the function $V_{\phi}/V$ in the \eqref{cossys} system near the point $\pi/2$:

\begin{equation}
\label{cossys2}
    \begin{aligned}
        \frac{d\phi}{dA} & =   X , \\
        \frac{dX}{dA} & = \bigg(\frac{X^2}{a^2} - 2 \bigg)\bigg(X + \frac{a^2}{\phi -\frac{\pi }{2}}-\frac{3+8a^2}{6} \left(\phi -\frac{\pi }{2}\right)\bigg).
    \end{aligned}
\end{equation}

Leaving the leading term in the system \eqref{cossys2} we are brought to
\begin{equation}
\label{cossys2-a}
    \begin{aligned}
        \frac{d\phi}{dA} & =   X , \\
        \frac{dX}{dA} & = \bigg(\frac{X^2}{a^2} - 2 \bigg)\frac{a^2}{\phi -\frac{\pi }{2}}. 
    \end{aligned}
\end{equation}
The system \eqref{cossys2-a} can be represented as a single  equation for the scalar field
\begin{equation}
\label{cossys2-b}
\left(\phi(A) - \frac{\pi}{2}\right) \phi''(A) - \left(\phi'(A)\right)^2 + 2 = 0 \,,
\end{equation}
where we define $\phi'(A) := d\phi/dA$ and set $a^2=1$.
Assuming  the following initial conditions
\begin{equation}
\label{incond}
    \phi(0) = \frac{\pi}{2} + \psi\,, \qquad \phi'(0) = \upsilon \,,
\end{equation}
where $\psi^2 \simeq 0$ and $-\sqrt{2}<\upsilon<\sqrt{2}$,
we get
the solution near the Minkowski point in the following form
\be\label{cossys2-c}
   \phi(A) = \frac{\pi}{2} + \frac{\psi}{\sqrt{
  2 - \upsilon^2}}\left(\sqrt{
  2 - \upsilon^2} \cos{\frac{A \sqrt{
  2 - \upsilon^2}}{\psi}} + 
   \upsilon \sin{\frac{A \sqrt{
  2 - \upsilon^2}}{\psi}}\right).
\ee

The solution \eqref{cossys2-c} is in agreement with the numerical solution of the system \eqref{cossys} under the initial conditions \eqref{incond}. This solution doesn't reach the Minkowski fixed point; however, it is worth noting that the equation \eqref{cossys2-b} is defined for the restricted region of $\phi$. Solutions to \eqref{cossys2-c} with different initial conditions form a family of nested ellipses with mutual vertices $(\phi,X)=(\frac{\pi}{2},\pm\sqrt{2})$ and with Minkowski point $M$ in their center. As we take into account the next nonlinear terms in \eqref{cossys2}, the solution to such a differential equation will be represented by the spiral curve covering these nested ellipses and eventually terminating at $\phi=\phi_{M}$. It is worth noting that an oscillatory approach to a Minkowski (FLRW) phase, originating from a de Sitter-like stage, was recently observed in \cite{Conzinu:2023fth}. In that work, the dilaton exhibits damped oscillations as it settles into the minimum of the effective potential after a regular bounce, leading to a final dust-dominated cosmology.


\subsubsection{Dynamical system in the disk $\mathbf{D}^2$ }
As it was mentioned above, the phase space of the dynamical system (\ref{cossys}) does not capture the behavior of the phase trajectories at infinity. However, we can improve our understanding of the dynamics near these regions if we map the system (\ref{cossys}) into the disk $\mathbf{D}^2$.

To perform this, we introduce  a new compact  coordinate $Z$ given by
\be\label{Zbigvar}
Z=2\tan\left(\frac{\phi}{2}\right).
\ee
By owning \eqref{Zbigvar}, we translate all periodic critical points to the finite set. 

 In general, to obtain a polynomial function of $Z$ for the term $V_{\phi}/V$, a suitable change can also be $Z=n\tan(\frac{\phi}{k})$, where $n,k\in\mathbb{N}$, and here we have chosen the most convenient case, $n=k=2$, which holds the structure  of the critical points on the phase portrait. One can also choose $k=1$, but in this case the points $\phi=\pm\frac{\pi}{2}$, which are related to Minkowski, correspond to $Z=\pm\infty$ and are not visible on the phase portrait. 

Next, we perform the  Poincaré transformation by introducing the following   change  of coordinates $(Z,X)\in\mathbf{R}^2\to (z,x)\in\mathbf{D}^2$:
\begin{equation} 
Z=\frac{z}{\sqrt{1-x^2-z^2}}, \quad
X=\frac{x}{\sqrt{1-x^2-z^2}},
\end{equation} 
with the constraint
\be
z^2+ x^2 \leq 1.
\ee
In the coordinates $(z,x)$ the autonomous dynamical \eqref{cossys} system takes the following form:
\begin{equation}\label{sysinD}
\begin{split}
\dot{z} = \mathrm{p}(z,x),\\
\dot{x} = \mathrm{q}(z,x),
\end{split}
\end{equation}
where the RHSs of equations are 
\begin{equation}
\begin{split}
\label{syscompdiskmain}
 \mathrm{p}(z,x)& = \frac{x}{4} \big(z^2-1\big) \left(4 x^2+3 z^2-4\right)-x z \left(3 x^2+2 z^2-2\right)\cdot\\ &\cdot \left(x+\frac{16 z
   \big(1-x^2-z^2\big) \big(48 (x^2-1) z^2+16 (x^2-1)^2+33 z^4\big)}{588 (x^2-1)
   z^4+368 (x^2-1)^2 z^2+64 (x^2-1)^3+285 z^6}\right),  \\
 \mathrm{q}(z,x)&=\frac{x^2(x^2-1)(z-4 x)}{4}+\big(1-x^2-z^2\big)\Biggl(\frac{z}{380} \big(1408-1693 x^2\big)+\big(x^2-1\big) x-\\&-\frac{4 z(7 x^4-9 x^2+2)}{5 (4 x^2+5 z^2-4)}-\frac{4z \big(x^2-1\big)
   \big(524 x^4+x^2 (591 z^2-820)-306 z^2+296\big)}{19 \big(72 (x^2-1) z^2+16
   (x^2-1)^2+57 z^4\big)}\Biggr).
   \end{split}
\end{equation}

\begin{figure}[H]
    \centering
\includegraphics[height=12cm]{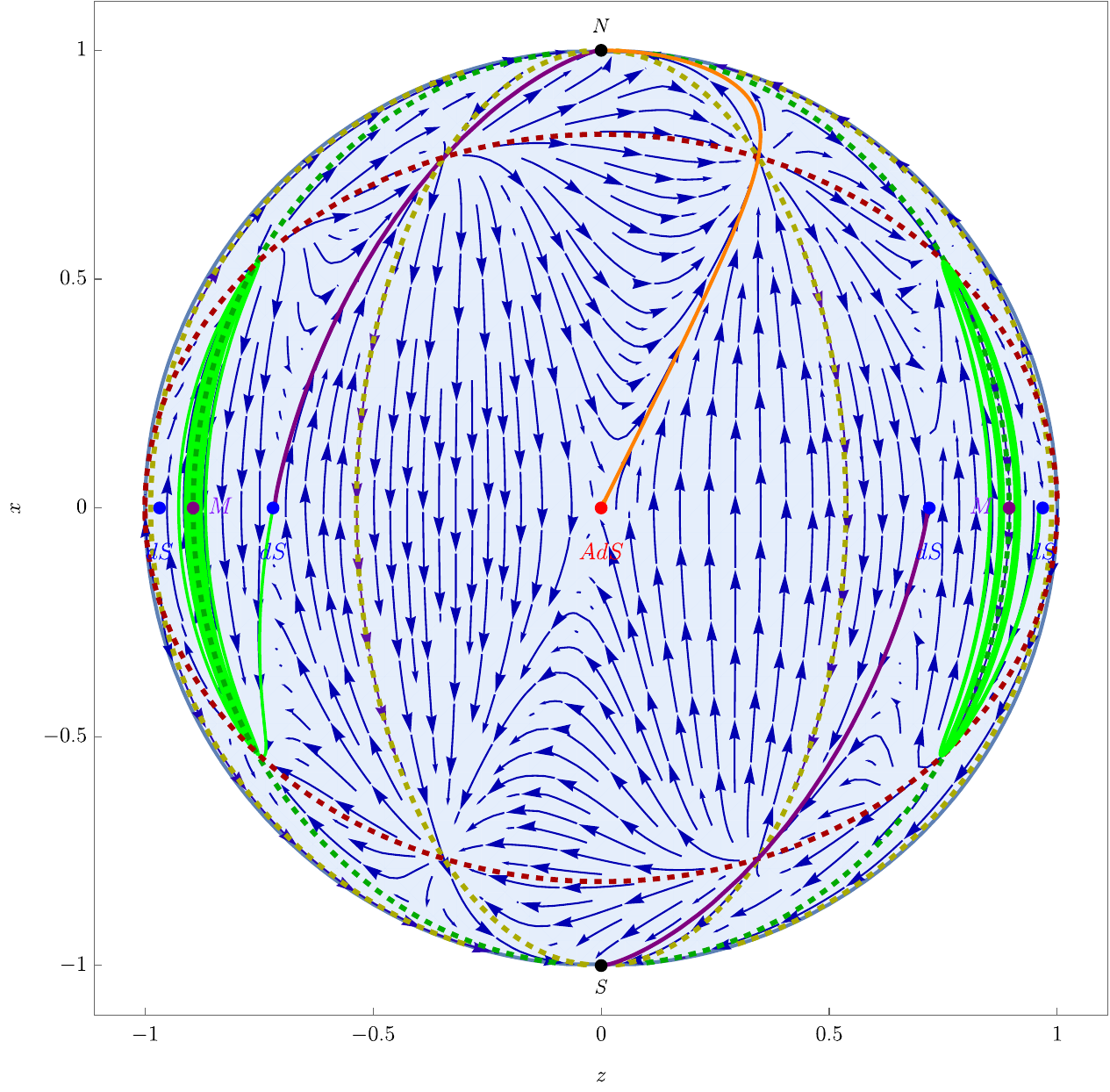}
    \caption{Global phase portrait for the system \eqref{sysinD} with \eqref{syscompdiskmain} in the  disk $\mathbb{D}^2$.}
    \label{fig:disktan}
\end{figure}

The critical points and sets of the system in the disk $\mathbf{D}^2$ (\ref{sysinD}) with (\ref{syscompdiskmain}) include
\begin{enumerate}
    \item $AdS$ : $(0,0)$, $dS$ : $(z_{dS},0)$, $N$ :  $(0,1)$, $S$ : $(0,-1)$;
    \item two curves corresponding to zeroes of the potential $V$, which are related to Minkowski extrema (shown by dashed green in Fig.~\ref{fig:disktan});
    \item four curves corresponding to the four zeroes of the potential except for zeroes related to Minkowski (shown by dashed dark yellow in Fig.~\ref{fig:disktan}).
\end{enumerate}

Note that $AdS$ and $dS$ critical points are saddles, while $N$ and $S$  are stable nodes. The phase portrait of the system in the disk \eqref{sysinD}  with \eqref{syscompdiskmain} is depicted in Fig.~\ref{fig:disktan}. Here we show the $AdS$ point by red, four $dS$ points by blue, two  Minkowski $M$ points by purple, and  points $N$ and $S$ corresponding to the poles of the disk $\mathbf{D}^2$ by black. Note that all Minkowski points correspond to each point of the singular green dashed curves, which, as the yellow dashed curves, are related to $V=0$. Here the latitudinal dashed red curves  correspond to the lines $X^2=2a^2$ of the system $\eqref{cossys}$ on the plane $\mathbf{R}^2$. The phase flows on the meridional curves are singular; however, on the intersections with the latitudinal dashed curves, they are regular. From Fig.~\ref{fig:disktan} we see that all flows pass these intersection points.  

 In Fig.~\ref{fig:disktan} we observe the following trajectories:
\begin{itemize}
    \item  the flow, which corresponds to the  exact  solution \eqref{Degsol}. This starts in $AdS$ and ends at the north pole $N$ of the disk, which corresponds to Minkowski.  In Fig.~\ref{fig:disktan}  we show this flow by an orange curve.  As mentioned above, this flow with Dirichlet b.c. is triggered by a non-zero VEV of an irrelevant operator with the scaling dimension $\Delta_{+}=2+2a^2$. 
    There is also a mirror flow from $AdS$ to the south pole $S$, corresponding to another branch of the solution \eqref{Degsol}. Note that  these exotic RG flows differ by the sign of the source;
     \item two numerical green flows, which begin in $dS$ and then go by spirals down to the points $M$ corresponding to Minkowski.
     The scaling dimension of the irrelevant operator is $\Delta_{+}=1+\sqrt{9+8a^2}$. 
     This flow previously appeared in Fig.~\ref{fig:cosflow} (in red); 
    \item the numerical purple flow, which also starts from the $dS$ point.  This flow we have already observed in Fig.~\ref{fig:cosflow}, and now it can be seen that it approaches Minkowski space in the same manner as the solution from AdS to Minkowski (the orange curve). The flow can be constructed  for any $dS$ point of the potential and can end either in the north pole $N$ or in the south pole $S$. This flow is related to an irrelevant operator with the same scaling dimension $\Delta_{+}$ as for the previous flow drawn by green.
\end{itemize}

\section{Warming up: flows with horizons}
\label{sec:3dflowsfneq1}
\subsection{Flows with $f\neq 1$ through 3d dynamical system}
In this section we will consider the case when we also have solutions with  the metric \eqref{genanmet} such that $f\neq 1$. This case includes an asymptotically AdS black hole (BTZ), a dS black hole and also a dS solution in the Poincar\'e coordinates. To describe such solutions, we introduce a new variable $Y$ related to $f$:
\begin{equation}
    Y=\frac{\dot{f}}{f\dot{A}},
\end{equation}
such that $Y\to\infty$ as $f\to 0$. 
Then EOM \eqref{eom1}-\eqref{eqd} are brought to the form of the autonomous 3d dynamical system defined on $\mathbf{R}^3$: 
\begin{equation}
\begin{split}
\label{3Dsyscos}
&\frac{d\phi}{dA}=X,\\ 
&\frac{dX}{dA}=\left(\frac{X^2}{a^2}-Y-2\right)\left(X+\frac{a^2}{2}\frac{V_{\phi}}{V} \right),\\ 
&\frac{dY}{dA}=Y\left(\frac{X^2}{a^2}-Y-2\right).
\end{split}
\end{equation}
In \eqref{3Dsyscos}, for any  fixed $\phi=const$, we have a critical curve given by
\be\label{cricurvfixedf}
\frac{X^2}{a^2}-Y-2=0,
\ee
similar to that one which appears for the autonomous dynamical system with the hyperbolic potential \cite{Golubtsova:2024dad}.

Since $Y\to \infty$, it is convenient to come to a compact configuration space as it was considered for the case of the hyperbolic scalar potential in \cite{Golubtsova:2024dad,Golubtsova:2024odp}. In this section, first, we will project the dynamical system \eqref{3Dsyscos} into the cylinder and, then, into the ball.
We will explore the space of solutions numerically and construct analytical solutions near the horizon.

\subsubsection{Dynamical system in the cylinder}

To study the flows, we apply the Poincaré transformation to map the dynamical system \eqref{3Dsyscos} into a unit cylinder as in \cite{Golubtsova:2024dad}. We keep the coordinate $\phi$ to be periodic, and the other two coordinates are transformed as
\begin{equation}\label{xysmall}
X=\frac{x}{\sqrt{1-x^2-y^2}} ,\quad
Y=\frac{y}{\sqrt{1-x^2-y^2}},
\end{equation}
such that the new coordinates $x$ and $y$ obey the constraint 
\be\label{xynewconstr}
x^2 + y^2\leq 1.
\ee
In these coordinates, the system \eqref{3Dsyscos} takes the following form:
\begin{equation}
\label{cyleqone}
    \begin{split}
\dot{\phi}& = \mathfrak{m}(x,y),\\
\dot{x} &= \mathfrak{p}(\phi,x,y),\\
\dot{y}& = \mathfrak{q}(\phi,x,y),
\end{split}
\end{equation}
where the r.h.s. of eqs.\eqref{cyleqone} are given by
\bea\label{mxy}
\mathfrak{m}(x,y)&=&x \sqrt{1-x^2-y^2};\\ \label{pxy}
\mathfrak{p}(\phi,x,y)&=&2 a^2 \left(x^2-1\right)\tan\phi\Big(\frac{ \left(x^2+y^2-1\right) \left(2 \sqrt{1-x^2-y^2}+y\right)}{(2 a^2+1) \cos 2\phi-2 a^2+1}+\nonumber\\
   &+&\frac{x^2\sqrt{1-x^2-y^2} \left(\left(2 a^2+1\right)\cos2\phi+1\right)}{(2 a^2+1)\cos2\phi-2a^2+1}\Big)+\nonumber\\
  &+&\frac{x \left(x^2+y^2-1\right) \left(a^2 \left(y \sqrt{1-x^2-y^2}+2(1 - x^2-
   y^2)\right)-x^2\right)}{a^2};\\ \label{qxy}
\mathfrak{q}(\phi,x,y)&=&   y \Big(2x \tan\phi\Big(\frac{a^2 (x^2+y^2-1) (2 \sqrt{1-x^2-y^2}+y)}{(2 a^2+1) \cos2\phi-2 a^2+1}+\nonumber\\&+&\frac{x^2
   \sqrt{1-x^2-y^2} ((2 a^2+1) \cos2\phi+1)}{(2 a^2+1) \cos2\phi-2 a^2+1}\Big)+\nonumber\\&+&\frac{(x^2+y^2-1) \left(a^2 (y \sqrt{1-x^2-y^2}+2(1-x^2-
   y^2))-x^2\right)}{a^2}\Big);
\eea
and we redefine the derivative as follows:
\begin{equation}
    \chi'=(1-x^2-y^2)d\chi/dA.
\end{equation}
This redefinition of the derivative is admissible (see  \cite{10.5555/102732}), since it does not change the behavior of flows and allows us to avoid divergencies that arise due to the constraint \eqref{xynewconstr}.

\begin{figure}[H]
    \centering
\includegraphics[height=5.5cm]{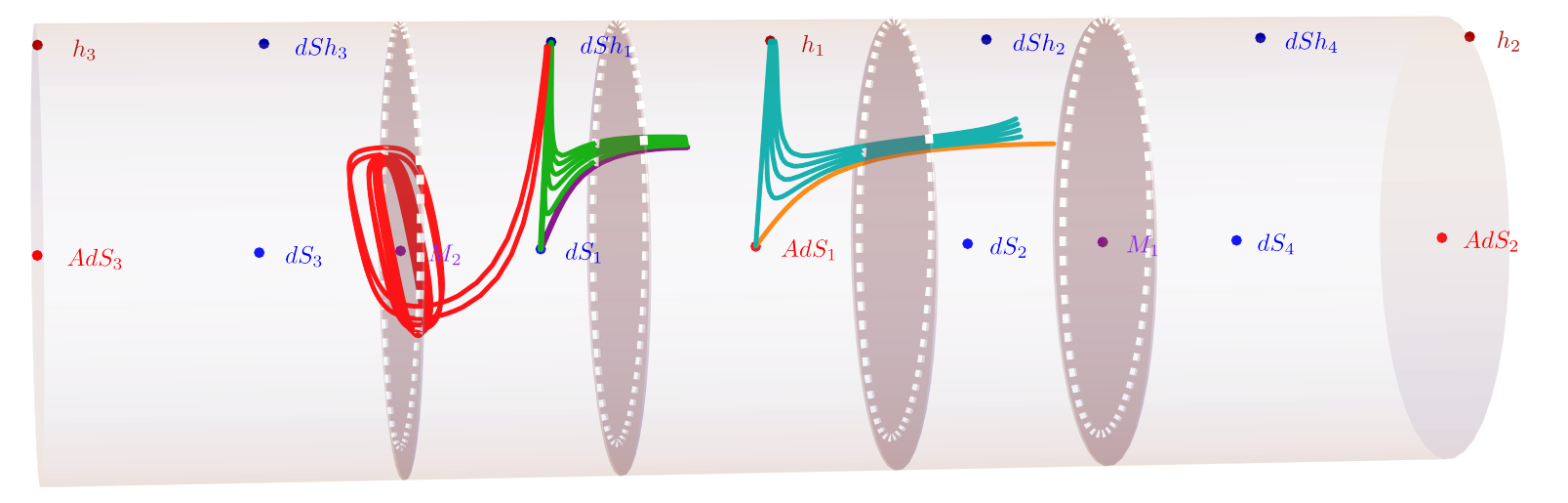}
    \caption{The trajectories of the dynamical system \eqref{cyleqone}, which phase space is a unit cylinder.
By the orange curve we show the exact solution from AdS$_{3}$ to Minkowski \eqref{Degsol} with $f=1$, 1(a). By straight cyan and green lines we show BTZ and SdS black holes, correspondingly 1(b) and 1(c). A family of solutions 2(a) is shown by cyan curves. The solutions 2(b) and 2(c) are shown by red and green curves, correspondingly. The solution 2(d) from dS to infinity is shown by purple. The singular surfaces are shown by gray, on which the white dashed curves are regular.}
    \label{fig:3dflowsspec}
\end{figure}

The phase portrait of the dynamical system \eqref{cyleqone} with \eqref{mxy}-\eqref{qxy} in period $\phi\in[-\pi,\pi]$ is presented in Fig.~\ref{fig:3dflowsspec}. The sets of the equilibrium points of \eqref{cyleqone} related to $f=0$ and $f=1$, correspondingly, are
\begin{enumerate}
    \item $AdS$: $(\phi_{AdS},0,0)$, $dS$: $(\phi_{dS},0,0)$; 
    \item $AdS_h$: $(\phi_{AdS},0,1)$, $dS_h$: $(\phi_{dS},0,1)$.
\end{enumerate}

The $\phi x$-plane in Fig.~\ref{fig:3dflowsspec} corresponds to the zero-temperature case, i.e., $f=1$, and is an invariant manifold of the system \eqref{cyleqone}. The curve \eqref{cricurvfixedf} in terms of the coordinates of the cylinder \eqref{xysmall} is shown by the white dashed closed curve.

Assuming certain initial conditions, we construct trajectories of \eqref{cyleqone}-\eqref{qxy} for $\phi\in(-\frac{\pi}{2},0]$, which we can see in Fig.~\ref{fig:3dflowsspec}, namely, 

\begin{enumerate}
\item the flows, which can be found exactly,

\begin{enumerate}
\item BTZ black hole with $f\neq1$, which starts from the horizon $h_{1}$ to the equilibrium  point, corresponding to AdS at $\phi=0$ as in \cite{Golubtsova:2024dad}. This solution is shown by the cyan straight line;
\item the Schwarzschild-de Sitter black hole with $f\neq1$, which flows from the horizon $dSh_{1}$ to  $dS_{1}$. This flow is shown by the green straight line;
\item the half-supersymmetric exact flow from AdS to Minkowski \eqref{Degsol} with $f=1$, shown by the orange solid curve. This trajectory also appears in Fig.~\ref{fig:disktan} (shown by orange);
\end{enumerate}
\item the flows, which are constructed numerically,
\begin{enumerate}
\item solutions that begin near the BTZ horizon $h_{1}$, pass close to $AdS$, and then go to infinity. These solutions are shown by cyan curves in Fig.~\ref{fig:3dflowsspec}. The initial conditions for such flows are $(\phi_{0},x_{0},y_{0})=(0+\delta,0,1-\epsilon)$. Here $0<\epsilon\ll1$ and $\delta\in(-|\phi_{n1}|,\phi_{n2})$, where $\phi_{n1}/\phi_{n2}$ corresponds to the null of the potential between $AdS$ and $dS_{1}/dS_{2}$ points;
\item solutions, which start near the SdS horizon $dSh_{i}$, pass near $dS$ and spiral down to Minkowski. These flows are shown by red curves in Fig.~\ref{fig:3dflowsspec}. The initial conditions for such flows are $(\phi_{0},x_{0},y_{0})=(\phi_{dS_{1}}-\delta,0,1-\epsilon)$. Here $0<\epsilon\ll1$ and $\delta\in(0,\frac{\pi}{2}-|\phi_{dS_{1}}|)$;
\item solutions, which start from SdS horizon $dSh_{i}$, pass near $dS$ and go to infinity. We show these solutions by green curves in Fig.~\ref{fig:3dflowsspec}. The initial conditions for such flows are $(\phi_{0},x_{0},y_{0})=(\phi_{dS_{1}}+\delta,0,1-\epsilon)$. Here $0<\epsilon\ll1$ and $\delta\in(0,|\phi_{n2}|)$, where $\phi_{n2}$ corresponds to the null of the potential between $AdS$ and $dS_{1}$ points;

\item a flow, which begins from the $dS$ point and behaves similarly to the exact solution \eqref{Degsol} with $f=1$. We show this flow by the purple curve in Fig.~\ref{fig:3dflowsspec}.  This trajectory also appears in Fig.~\ref{fig:disktan} (in purple).
\end{enumerate}
\end{enumerate}
\begin{table}[ht]
    \centering
    \begin{tabular}{|c||c|c|c|c|}
    \hline
                points set&  type & eigenvalues ($-\Delta_{\phi},-\Delta_{x},-\Delta_{y}$)  \\
    \hline
    \hline
         $(\phi_{0},0,1)$  & unstable degenerate node  & $(0,0,2)$  \\
         \hline
       $(\phi_{M},1,0)$ & stable degenerate node    &  $(-2,0,0)$ \\
       \hline
       $(\phi_{AdS},0,0)$    & saddle & $(-4,2,-2)$\\
       \hline
       $(\phi_{dS},0,0)$    & saddle& $(-1-\sqrt{17},-1+\sqrt{17},-2)$\\
       \hline 
    \end{tabular}
    \caption{Classification of the critical points of the dynamical system \eqref{cyleqone} (we set $a^2=1$). Here $\phi_{0}\in[-\pi,\pi]$ and we suppose that $(\phi_{0},0,1)$ represents the set of points on the cylinder axis with $y=1$.}
    \label{tab:typescyl}
\end{table}

We see that solutions 2(a)-2(d) cross the singular surface through the dashed white curves.  

From Fig.~\ref{fig:3dflowsspec} we observe that the BTZ and SdS black holes are the only black holes with AdS/dS asymptotics for the system \eqref{cyleqone}. In that sense, we cannot construct a family of asymptotically AdS/dS black hole solutions that are characterized by some value of the scalar field on the horizon as it was done for the hyperbolic potential \cite{Golubtsova:2024dad}. Instead, a small variation of the scalar field $\phi$ leads to solutions that begin near horizon regions of BTZ/SdS black holes, then pass near $AdS/dS$ critical points, and then go to infinity. For the AdS case, this can be interpreted as that for exotic RG flows driven by VEV of irrelevant operators,  there is only a  thermal state, which is described by the BTZ black hole. 

Below we will also discuss the system with $f\neq 1$ projected into $\mathbb{B}^3$.

\subsubsection{Dynamical system in the ball}

In this subsection
to clarify the behavior of the flows, which we discussed above,
we will map the 3d dynamical system \eqref{3Dsyscos} into the ball. For this we use a compact variable $Z$ \eqref{Zbigvar} instead of the periodic $\phi$ in \eqref{3Dsyscos}, i.e., we get the following system:
\begin{equation}
\begin{split}
\label{3DsysXZY}
&\frac{dZ}{dA}=X\left(1+\frac{Z^2}{4}\right),\\ 
&\frac{dX}{dA}=\left(\frac{X^2}{a^2}-Y-2\right)\left(X+\frac{a^2}{2}\frac{V_{\phi}}{V} \right),\\ 
&\frac{dY}{dA}=Y\left(\frac{X^2}{a^2}-Y-2\right).
\end{split}
\end{equation}
Then we apply the Poincaré projection to map the dynamical system \eqref{3DsysXZY} defined in $\mathbb{R}^3$ into the ball $\mathbb{B}^3$, namely,
\begin{equation}
\label{coordball}
        Z=\frac{z}{\sqrt{1-z^2-x^2-y^2}},\quad
         X=\frac{x}{\sqrt{1-z^2-x^2-y^2}},\quad
          Y=\frac{y}{\sqrt{1-z^2-x^2-y^2}},
\end{equation}
where  $(z,x,y)\in \mathbb{B}^3$ and are also related by the constraint
\begin{equation}
    x^2+y^2+z^2\leq 1.
\end{equation}

The coordinate transformation \eqref{coordball} allows us to map the infinite points of the system $\eqref{3Dsyscos}$, which, as we have seen, are difficult to be caught studying the dynamical system in the unit cylinder, into the ball.
Then the dynamical system in the ball takes the form
\begin{equation}
\label{cyleball}
    \begin{split}
\dot{z}& = \mathrm{m}(z,x,y),\\
\dot{x} &= \mathrm{p}(z,x,y),\\
\dot{y}& = \mathrm{q}(z,x,y),
\end{split}
\end{equation}
where the r.h.s. of \eqref{cyleball} can be found from \eqref{3DsysXZY} applying \eqref{coordball}.

\begin{figure}[H]
    \centering
\includegraphics[height=10cm]{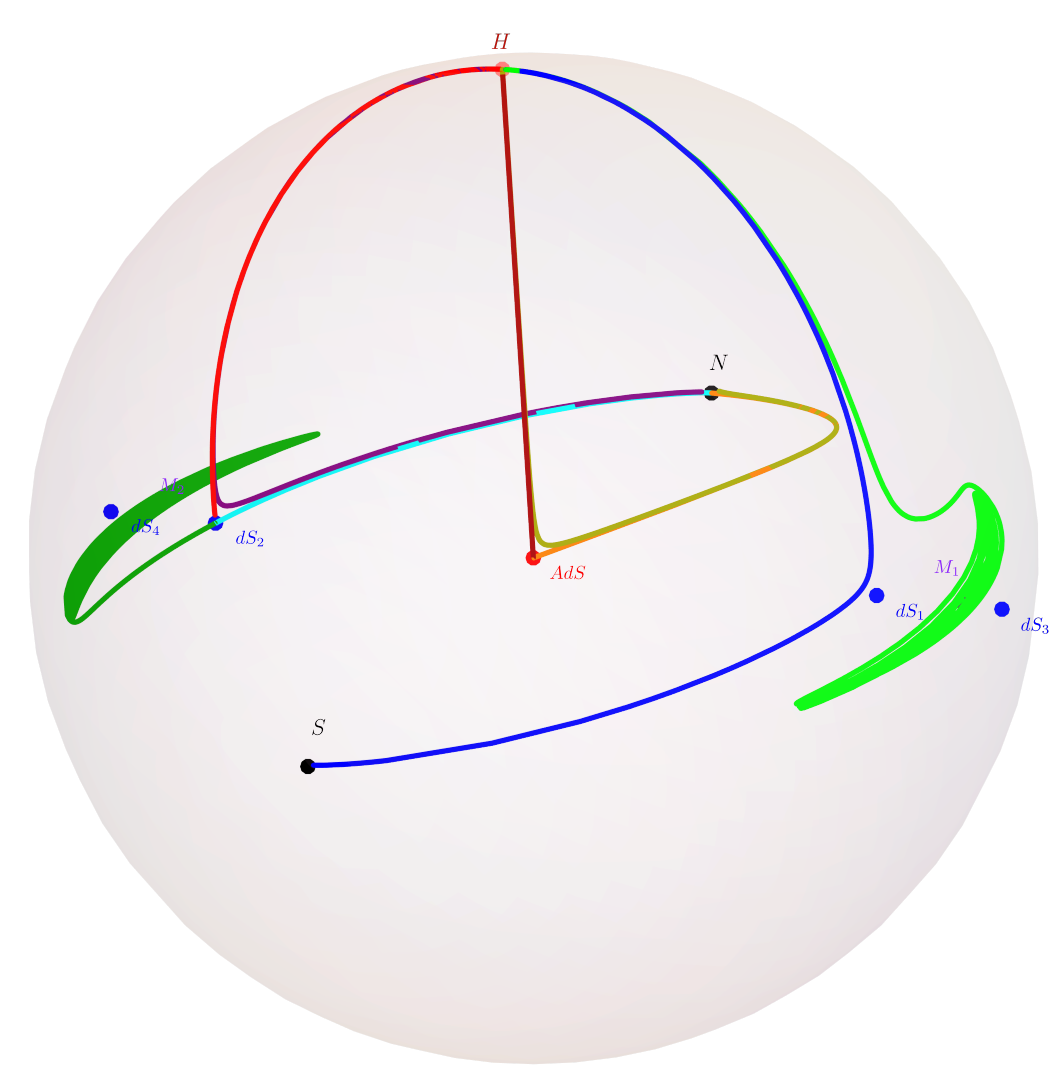}
    \caption{The numerical flows of the system \eqref{3Dsyscos}, which phase space is a 3d ball. 
    The straight red line from $H$ to $AdS$ is the BTZ solution. The solid red curve from $H$ to $dS_2$ describes the Schwarzschild-de Sitter solution. The dark yellow curve from $H$ to $N$ corresponds to a singular deformation of the BTZ solution. The blue and purple curves from $H$ to $N$ and $S$, correspondingly, are related to deformations of the SdS solution. The green curve from $H$, which spirals down to Minkowski, also corresponds to a deformation of the SdS solution.
    The orange curve from $AdS$ to $N$ is the exact zero-temperature flow. The solution from $N$ (Minkowski) to dS is shown by cyan. The dark green curve corresponds to a zero-temperature flow from dS to Minkowski.}
    \label{fig:RGballflows}
\end{figure}

Under the coordinate transformation \eqref{coordball}, the points of the system on $\mathbb{{R}}^3$ with the coordinate $Y\to+\infty$ are mapped to the point $(z,x,y)\to(0,0,1)$ in the ball $\mathbb{B}^3$, regardless of the values of $Z$ and $X$ if they are fixed. That is, all the points in the cylinder with coordinates $(\phi,0,1)$ for any $\phi$ in Fig.~\ref{fig:3dflowsspec} are mapped into the point $(z,x,y)=(0,0,1)$ in $\mathbb{B}^3$. We denote this point by $H$ in Fig.~\ref{fig:RGballflows} as corresponding to horizons of both BTZ and Schwarzschild-de Sitter solutions. The types of critical points are considered in Table \ref{tab:typesball}.

We show the phase portrait of the system in the unit ball \eqref{cyleball} in Fig.~\ref{fig:RGballflows}.
The center of the ball $(0,0,0)$ corresponds to the AdS solution at zero temperature. The dS points are inside the sphere and have coordinates $(z_{dS},0,0)$. The Minkowski points with $\phi=\pm\pi/2$ are mapped  to the critical curve, which includes the north and the south poles, $N$ and $S$, with the coordinates $(0,\pm1,0)$ and points $M$. 
 All trajectories corresponding  to the finite temperature solutions start from the horizon $H$ with  $(0,0,1)$.

\begin{table}[ht]
    \centering
    \begin{tabular}{|c||c|c|c|c|}
    \hline
                point&  type & eigenvalues ($-\Delta_{z},-\Delta_{x},-\Delta_{y}$)  \\
    \hline
    \hline
         $H$  & unstable node  & $(1,0,2)$  \\
         \hline
       $N,S$&  stable node     &  $(-2,-1,0)$ \\
       \hline
       $AdS$    & saddle & $(-4,2,-2)$\\
       \hline
       $dS$    & saddle& $(-1-\sqrt{17},-1+\sqrt{17},-2)$\\
       \hline 
    \end{tabular}
    \caption{Classification of critical points of the dynamical system \eqref{cyleball} for $a^2=1$.   }
    \label{tab:typesball}
\end{table}

We see the following flows to \eqref{cyleball} in  Fig.~\ref{fig:RGballflows} are:

\begin{enumerate}
\item flows that can be found exactly,
\begin{enumerate}
    \item the exact non-rotating BTZ solution, shown by the straight dark red line going from $H$ to $AdS$;
   \item the exact SdS solution shown by the red curve from $H$ to $dS_{2}$;
   \item the exact supersymmetric flow with $f=1$ shown by the orange curve from $AdS$ to $N$.
\end{enumerate}
\item flows, which can be constructed numerically,
\begin{enumerate}
\item the solution, which flows from $H$, passes the $AdS$ point and ends in the north pole $N$. We show this by the solid dark yellow curve. To construct the solution, we take the initial conditions as $(z_{0},x_{0},y_{0})=(0,\delta,1-\epsilon)$, where $\delta,\epsilon>0$ and $\delta,\epsilon\ll 1$. This solution appears as a deformation of the BTZ black hole, introducing a non-zero finite value of $\phi$ on the horizon;

\item the flows from $H$, passing near $dS$ and ending in $N/S$ points, respectively. The flows are shown by purple and blue curves in Fig.~\ref{fig:RGballflows}. To find them we use the following initial conditions: $(z_{0},x_{0},y_{0})=(\pm\varepsilon,\pm\delta,1-\epsilon)$, where $\delta,\epsilon,\varepsilon>0$ and $\delta,\epsilon,\varepsilon\ll 1$.  These flows correspond to deformations of the SdS solution such that on the horizon we have $\phi(w_h)=\phi_h$;

\item the trajectory shown by green corresponds to a deformation of the SdS solution and flows from $H$, passing near $dS_{1}$, then spiraling down to the Minkowski point $M$.  The initial conditions for this flow read as $(z_{0},x_{0},y_{0})=(\pm\varepsilon,\mp\delta,1-\epsilon)$, where $\delta,\epsilon,\varepsilon>0$ and $\delta,\epsilon,\varepsilon\ll 1$;
\item the flows with $f=1$, 
previously discussed in Sec.~\ref{sec:2dflowsf1}, see in Fig.~\ref{fig:disktan}, namely, the flow from $dS_{2}$ to $N$ by cyan and the flow from $dS_2$ to Minkowski by dark green.

\end{enumerate}
\end{enumerate}

The trajectories of the dynamical system \eqref{cyleball}
in Fig.~\ref{fig:RGballflows} 
can clarify the behavior of the flows from Fig.~\ref{fig:3dflowsspec}.
Namely, we see that the flows 2(a) and 2(b), which correspond to deformations of the BTZ and SdS solutions, start from $H$, which is an unstable  node, and end in the north pole $N$, which is a stable node, see Table~\ref{tab:typesball}. We see that these deformations end in  Minkowski.

It is instructive  to do the Lyapunov analysis of the dynamical system \eqref{cyleball}. The eigenvectors at the critical point $H$, which is an unstable node, are given by:
\be
\label{3deigenvec}
u_{z} =\left[\begin{array}{c}
1\\
0\\0\end{array}\right],\quad u_{x} =\left[\begin{array}{c}
1\\\frac{1}{4}\\
0\end{array}\right], \quad u_{y} =\left[\begin{array}{c}
0\\0\\
1\end{array}\right].
\ee
Because the one eigenvalue is equal to zero, the solution near the point $H$ is not linear. Though, keeping in mind the form of eigenvectors \eqref{3deigenvec}, we can schematically write down the solution near the point $H$ 
\bea
\label{Luapball}
\left[\begin{array}{c}
z\\
x\\
y-1
\end{array}\right]=\left[\begin{array}{c}
1\\
0\\0\end{array}\right]\big(c_{1z}+c_{2z}N(A)\big)e^{-\Delta_{z}A}+\left[\begin{array}{c}
0\\
1\\0\end{array}\right]\big(c_{1x}+c_{2x}N(A)\big)+\left[\begin{array}{c}
0\\
0\\1\end{array}\right]c_{y}e^{-\Delta_{y}A},
\eea
where $c_{1z}$, $c_{2z}$, $c_{1x}$, $c_{2x}$ and $c_{y}$ are non-zero constants characterizing the solutions. Here by $N(A)$ we denote the non-linear and logarithmic terms,  which contribute to the solution of the scalar field $\phi$. We see the correspondence between constants and solutions in Table \ref{tab:typesballsol}.

\begin{table}[ht]
    \centering
    \begin{tabular}{|c|c|c|c|c|}
    \hline
                type&  solution & zero constants &non-zero constants  \\
    \hline
    \hline
         1(a)  & BTZ  & $c_{1z}, c_{2z}, c_{1x}, c_{2x}$ & $c_{y}$ \\
         \hline
       1(b)&  SdS     &  $c_{2z}, c_{1x}, c_{2x}$ & $c_{1z}, c_{y}$\\
       \hline
       2(a)    & deformed BTZ & $c_{1z}, c_{2z}$ & $c_{1x}, c_{2x}, c_{y}$\\
       \hline
       2(b), 2(c)    & deformed SdS & - & $c_{1z}, c_{2z}, c_{1x}, c_{2x}, c_{y}$\\
       \hline 
    \end{tabular}
    \caption{Types of solutions depending on the constants in schematic Lyapunov solution \eqref{Luapball}.   }
    \label{tab:typesballsol}
\end{table}

In the next section,  we will  show that the scalar field near the horizon of the singular solutions 2(a)-2(c) contains the logarithmic divergences, such behavior of the scalar field  allows us to bring the flows to Minkowski. 

\subsection{Analytical solutions}
\label{sec:analytsol}

\subsubsection{Exact solutions with horizons}


First, we recall  how to derive  dS$_{3}$, AdS$_{3}$, 
BTZ and Schwarzschild-de Sitter solutions  from the supergravity equations. All of them correspond to extrema of the  potential. 
Assuming that a solution  is defined for $V'(\phi)=0$, the equation for the scale factor is easily integrated
\be
\label{AC1C2}
A=C_{1}w+C_{2},\quad C_{1},C_{2}=\text{const},
\ee
where  $C_2$ can be set to zero. A non-zero $C_1$ corresponds to AdS and dS solutions, while $C_1=0$ is valid for the Minkowski solution.  Then, we plug \eqref{AC1C2} into \eqref{eom2}, which yields
\be\label{adsdses}
f= -\frac{2V}{C_{1}^2}+\bar{c}e^{-2C_{1}w}, \quad \bar{c}=\text{const}.
\ee
From the latter, one can see that the dS$_3$ metric (\ref{dSfcoord}) corresponds to the scale factor \eqref{adsdses} with $C_{1}=\sqrt{2V}$,  $V>0$ and $\bar{c}=0$ .
Note that in terms of the coordinates of the dynamical system on $\mathbb{R}^3$ \eqref{3Dsyscos}, the  dS$_3$ equilibrium point is given by $X=0,  Y=0, \phi=\phi_{dS}$. 
If $V<0$, $C_{1}=\sqrt{-2V}$, and $\bar{c}=0$, we get the AdS$_{3}$ metric. Taking $\bar{c}$ to be  non-zero, $\bar{c}\neq0$, we get AdS and dS black hole solutions, namely, the blackening function for the BTZ/SdS$_3$ solutions reads
\be
f(w-w_{h})=\pm 1\mp e^{-2\sqrt{\mp 2V}(w-w_h)},
\ee
where we fixed $\bar{c}=\mp 1$ for the BTZ and SdS solutions, correspondingly.
The scale factor of BTZ/SdS black hole is given by \eqref{AC1C2}.

Also note that for both BTZ and SdS$_{3}$ solutions the coordinate $Y$ has the same sign of infinity in the limit $w\to w_{h}$:
\be
Y=\frac{\dot{f}}{f\dot{A}}\underset{w\to w_{h}}{\to}+\infty, \quad \dot{A}=C_{1}=\text{const},
\ee
hence we can see both BTZ and SdS$_{3}$ solutions on the same phase portrait in Fig.~$\ref{fig:RGballflows}$.

\subsubsection{Analytical solutions near horizon}

 To find an analytical description of the numerically constructed solutions from the previous subsection near the horizon regions, we follow the method from \cite{Golubtsova:2024dad,Golubtsova:2024odp}. In the  near-horizon region, $Y\to+\infty$, the system \eqref{3Dsyscos} simplifies, so the equation for $Y$ decouples from the system and can be easily solved as
\begin{equation}\label{Yinf}
    Y(A-A_{h})=
    \frac{1}{A-A_{h}},
\end{equation}
where the integration constant was chosen to properly obtain the boundary condition $Y(A_{h})=+\infty$. 
The other two equations of the system \eqref{3Dsyscos} are reduced to a single second-order equation
for the scalar field \begin{equation}\label{phionA}
    \big(A-A_{h}\big)\frac{d^2\phi}{dA^2}+\frac{d\phi}{dA}+\frac{a^2}{2}\frac{V'(\phi)}{V(\phi)}=0,
\end{equation}
where we also used \eqref{Yinf}.

We are looking for near-horizon geometries when the value of the scalar field at the horizon $\phi_{h}$ is not necessarily the value at the extremum of $V(\phi)$. We can further simplify eq. \eqref{phionA} by expanding the last term in a Taylor series up to the first order near $\phi_h$. As a result, we obtain the following equation:
\begin{equation}
\label{eqonphihor1}
    \mathcal{A}\frac{d^2\Phi}{d\mathcal{A}^2}+\frac{d\Phi}{d\mathcal{A}}+\Delta_{h}\Phi+\Lambda_{h}=0,
\end{equation}
where we use  $\Phi=\phi-\phi_{h}$ and $\mathcal{A}=A-A_{h}$. In \eqref{eqonphihor1}, we also denote the first two terms in the Taylor expansion of $V'/V$ as
\begin{equation}\label{deltah}
    \Delta_{h}=\frac{a^2}{2}\left(\frac{V''(\phi_{h})}{V(\phi_{h})}-\left(\frac{V'(\phi_{h})}{V(\phi_{h})}\right)^2\right), \quad \Lambda_{h}=\frac{a^2}{2}\frac{V'(\phi_{h})}{V(\phi_{h})}.
\end{equation}
One can see that for any $\phi_{h}$ and $a^2$, the quantity $\Delta_{h}$ in \eqref{deltah} is negative 
\begin{equation}\label{deltahgen}
    \Delta_{h}=-4a^2\frac{1-4a^4+(1+2a^2)\cos2\phi_{h}+4a^4\sec^2\phi_{h}}{\big(1-2a^2+(1+2a^2)\cos2\phi_{h}\big)^2}<0.
\end{equation}

Then, taking into account \eqref{deltahgen}, the general solution to \eqref{eqonphihor1} can be represented as follows: 

\begin{equation}
\label{gensolhor1}
    \phi(A)=\phi_{h}-\frac{\Lambda_{h}}{\Delta_{h}}+
\mathrm{c_1} I_0\left(2 \sqrt{|\Delta_{h}|(A-A_{h})}\right)+ \mathrm{c_2} K_0\left(2 \sqrt{|\Delta_{h}|(A-A_{h}) }\right),
\end{equation}
where  $I_{0}$ and $K_{0}$ are modified Bessel functions of the first and second kinds, respectively.

Imposing the condition  on the horizon 
\be
\label{bcphi1}
\phi(A_h)=\phi_h + \delta,
\ee
with $\delta^2 \sim 0$ and expanding in series the solution \eqref{gensolhor1} near $A_h$, we get
\begin{equation}
\label{approxphisol}
    \phi=\phi_{h}+\frac{\Lambda_{h}}{|\Delta_{h}|}+\mathrm{c_1}\big(1+|\Delta_{h}|\bar{A}\big)+\mathrm{c_2}\Big(-2\gamma-\log\bar{A}+\big(2-2\gamma+\log\bar{A}\big)\bar{A}\Big),
\end{equation}
where $\bar{A}=A-A_{h}$. Setting the integration constant $\mathrm{c}_2=0$, we neglect the contribution of $K_0$ to the solution $\eqref{gensolhor1}$, which is divergent on the horizon. In this case we obtain only solutions 1(a) and 1(b) in Fig. \ref{fig:RGballflows}. For them the constant $\mathrm{c}_1$ is given by
\begin{equation}
\label{constc1hor}
\mathrm{c_1}=\delta-\frac{\Lambda_{h}}{|\Delta_{h}|}.
\end{equation}
To obtain the solutions 2(a)-2(c) in Fig. \ref{fig:RGballflows}, we should also include $K_{0}$ and for the initial conditions impose $\phi'(A_{h})=\upsilon$ together with \eqref{bcphi1}. The integration constants in this case become
\begin{equation}
\begin{split}
    &\mathrm{c_1}=\frac{\big(\delta-\frac{\Lambda_{h}}{|\Delta_{h}|}\big)\left(1+\big(-3+2\gamma+\log(|\Delta_{h}|\epsilon)\big)|\Delta_{h}|\epsilon\right)-\big(2\gamma+\log(|\Delta_{h}|\epsilon)\big)\epsilon v}{1-2|\Delta_{h}|\epsilon\big(1+\log(|\Delta_{h}|\epsilon)\big)},\\
 &\mathrm{c_2} =-\frac{\big(\Lambda_{h}-|\Delta_{h}|\delta+v\big)\epsilon}{1-2|\Delta_{h}|\epsilon\big(1+\log(|\Delta_{h}|\epsilon)\big)},
 \end{split}
\end{equation}
where we fix $\epsilon=(A-A_{h})$ to be small and neglect terms of order $\epsilon^2$. Near  the horizon, $\epsilon\to0$, the constant $c_{1}$ resembles \eqref{constc1hor}, and the constant $c_{2}$ goes to zero. However,  the term with $c_{2}$ grows as we move away from the horizon, causing the solution to deviate from the regular BTZ/SdS geometry and become singular.

For the case $c_{2}=0$ and  $c_{1}$ as in \eqref{constc1hor} we can easily write down an analytical solution near the horizon. Having \eqref{approxphisol}, the equation on the scale factor $A(w)$ \eqref{eom1} leads to the solution
\begin{equation}\label{scaleA}
    A(w)=\Theta_{h}\ln\left(1+C_{A}\Theta_{h}^{-1}w\right),\quad \Lambda^{'}_{h}=\delta|\Delta_{h}|-\Lambda_{h},\quad \Theta_{h}=\frac{a^2}{\Lambda^{'2}_{h}},
\end{equation}
where the constant $C_{A}=\sqrt{|2V(\phi_{h})|}$ is fixed such that in the extremum of the potential the solution \eqref{scaleA} reproduces the BTZ/SdS black hole solutions near the horizon.

As in \cite{Golubtsova:2024dad,Golubtsova:2024odp}, we have for the blackening function $f(A)\approx C_{f}(A-A_{h})$, so to find the constant $C_{f}$ we plug \eqref{approxphisol} and \eqref{scaleA} into  \eqref{eom3}, which yields
\begin{equation}
    \dot{A}^2\left(\frac{1}{A-A_{h}}+2-\Theta_{h}^{-1}\right)+\frac{4V(\phi_{h})}{C_{f}(A-A_{h})}=0.
\end{equation}
Neglecting the second and the third terms in parentheses in the latter equation, we get $C_{f}$:
\be
    C_{f}=2 \, \text{sgn}\big(V(\phi_{h})\big)\Bigl(1+C_{A}\Theta_{h}^{-1}w_{h}\Bigr)^2,
\ee    
such that near extrema, where $\Lambda^{'}_h\to 0$, we have
\be\label{Cfadsds}
C_{f} =2\,\text{sgn}\big(V(\phi_{AdS/dS})\big).
\ee
We see that the sign of $C_{f}$ depends on the value of $V(\phi_{h})$, i.e., for the trajectories on the interval $\phi_{h}\in(-\frac{\pi}{2},0]$ in the Fig.~\ref{fig:3dflowsspec} it is negative for $\phi_{h}\in(\frac{1}{2}\arccos\left(\frac{1-2a^2}{1+2a^2}\right),0]$ and positive for $\phi_{h}\in(-\frac{\pi}{2},\frac{1}{2}\arccos\left(\frac{1-2a^2}{1+2a^2}\right))$.

Then the metric of the solution in the near horizon region looks like
\be\label{metricnh}
ds^2= \left(1+C_{A}\Theta_{h}^{-1}w\right)^{2\Theta_{h}}\left(-f(w)dt^2+ dx^2\right) +\frac{dw^2}{f(w)},
\ee
with blackening function
\begin{equation}\label{blackfnh}
    f(w)=C_{f}\Theta_{h}\ln\left(\frac{1+C_{A}\Theta_{h}^{-1}w}{1+C_{A}\Theta_{h}^{-1}w_{h}}\right),
\end{equation}
and the scalar field near the horizon is given by
\be\label{phinh}
\phi = \phi_{h}+\delta + \frac{a^2}{\Lambda^{'}_{h}}\ln\left(\frac{1+C_{A}\Theta_{h}^{-1}w}{1+C_{A}\Theta_{h}^{-1}w_{h}}\right),
\ee
with the constants $C_{f},C_{A}$, $\Lambda_{h}^{'}$ and $\Theta_{h}$ defined above. 

If the horizon is located in the extrema of the potential, such that $\Lambda_h^{'}\to 0$, the solution \eqref{metricnh} come to the near-horizon geometry of the BTZ/SdS black hole. The metric then turns into the form
\be
ds^2 \approx e^{2C_A w_h}\left(-f(w)dt^2 + dx^2\right) + \frac{dw^2}{f(w)},
\ee
with
\be
\label{fsolbtzsds}
    f(w)
    \approx C_{f}C_{A}(w-w_{h}),
\ee
where $C_f$ is given by \eqref{Cfadsds} and $C_{A}=\sqrt{|2V(\phi_{AdS/dS})|}$ for the AdS/dS case, i.e.,
\begin{equation}
\begin{split}
    \quad C_{f}=2, \quad C_{A}=\sqrt{-2V(\phi_{AdS})}\quad \text{for} \quad V(\phi_{AdS})<0;\\
    \quad C_{f}=-2, \quad C_{A}=\sqrt{2V(\phi_{dS})}\quad \text{for} \quad V(\phi_{dS})>0.
\end{split}
\end{equation}

In \cite{Golubtsova:2024dad,Golubtsova:2024odp}, a special class of the flows with a slowly changing scalar field $X^2\sim0$ was found that almost match with the flows without the constraint.  The geometry of the flows is described by a non-deformed BTZ metric, and the solution to the scalar field is given by a hypergeometric function, which can be extended from the near-horizon region to the boundary.

Under the constraint $X^2\sim 0$, the solution to $Y$ is  $Y(A-A_{h})=\frac{2}{e^{2(A-A_{h})}-1}$. 
Then, substituting $Y(A)$ back into the system \eqref{3Dsyscos}, we derive the equation for the scalar field
\begin{equation}
    r\big(1-r\big)\frac{d^2\Phi(r)}{dr^2}+\big(1-2r\big)\frac{d\Phi(r)}{dr}+\frac{|\Delta_{h}|}{2}\Phi(r)=0,
\end{equation}
where $r=e^{2(A-A_{h})}$ and $\Phi=\phi-\phi_{h}+\frac{\Lambda_{h}}{\Delta_{h}}$.
The fundamental solution to this equation near $r=1$ can be represented in the form
\begin{equation}
    \Phi(r)={}_{2}F_{1}(a_{h},1-a_{h},1,1-r),
\end{equation}
where $a_{h}=\frac{1}{2}\big(1+\sqrt{1+2|\Delta_{h}|}\big)\geq1$, which means this solution diverges for $r\to\infty$. We see that unlike in \cite{Golubtsova:2024dad}, such a solution is not able to be analytically defined from horizon to the conformal boundary.

\section{Exact RG flow and "$T\overline{T}$" operator}
\label{Sect:5}

Let us calculate the Brown-York stress-energy tensor of the exact holographic RG flow given by \eqref{Degsol} with $f = 1$, $a^2 = 1$, $m = 1$ assuming a cutoff surface $\Sigma_c$ at $w = w_c$. 
For the metric \eqref{Degsol}, the unit normal to the hypersurface $\Sigma_c$ is $\partial_w$ with the components
\be 
n^{\mu}=(0,0,1),
\ee
then the induced metric reads
\be\label{indmetricwc}
\gamma_{ab}dx^a dx^b = e^{2A(w_c)}\big(-dt^2 + dx^2\big).
\ee
Taking into account \eqref{indmetricwc}, the components of the extrinsic curvature are defined as
\be
K_{ab} = \frac{1}{2}\mathcal{L}_n \gamma_{ab} = \frac{1}{2}\partial_w \gamma_{ab}
\ee
and we have
\be
K_{tt}=-K_{xx} 
= -A'(w)e^{2A(w)},\quad
K
= 2A'(w).
\ee
The Brown-York tensor with  the counterterm is given by
\be\label{BrownYork}
T^{BY}_{ab} = \frac{1}{8\pi G} \left(K_{ab} - K\gamma_{ab} + \frac{1}{\ell}\gamma_{ab}\right),
\ee
where $\ell = \frac{1}{2}$ from \eqref{massads}. The non-zero components  of \eqref{BrownYork} at $w = w_c$ are
\be\label{nonzeroBY}
T^{BY}_{tt} 
=-T^{BY}_{xx}= \frac{1}{8\pi G} e^{2A} \big(A' - 2\big). 
\ee

One can rewrite the Brown-York tensor at $w = w_c$  \eqref{nonzeroBY} in terms of  
the coupling constant $\phi$
\be\label{SETBYAB}
T^{BY}_{ab} = \frac{e^{2A(w)}}{8\pi G} \begin{pmatrix} \frac{2}{1 + e^{8w}} - 2 & 0 \\ 0 & -\frac{2}{1 + e^{8w}} + 2\end{pmatrix} = \frac{e^{2A(\phi)}}{4\pi G} \begin{pmatrix} -\sin^2 \phi & 0 \\ 0 & \sin^2 \phi \end{pmatrix},
\ee
where we used that
\be
A'(w) =
\frac{2}{1 + e^{8w}}=2\cos^2 \phi.
\ee
Next, we can compute the trace of the Brown-York stress-energy tensor\footnote{For readability, we now omit the superscript BY.}
\be\label{traceset}
T_{a}^{a} =\frac{1}{2\pi G}\sin^2 \phi
\ee
and the square of the stress-energy tensor
\bea\label{TTabab}
\begin{split}
T^{ab}T_{ab} =\gamma^{ac}\gamma^{bd}T_{ab}T_{cd} 
=\frac{1}{8\pi^2 G^2}\sin^4 \phi.
\end{split}
\eea
Taking into account \eqref{traceset}-\eqref{TTabab}, we can define the composite $T\overline{T}$ operator:
\be\label{ToverT}
T\overline{T}=\frac{1}{8}\big(T^{ab}T_{ab} - (T^a_a)^2\big) 
= -\frac{1}{64\pi^2 G^2}\sin^4 \phi.
\ee
For any stationary, translation‑invariant state in any 2d QFT, the expectation value of the $T\overline{T}$ factorizes as $\Braket{T\overline{T}}=\Braket{T}\displaystyle{\braket{\overline{T}}}-\Braket{\Theta}^2$. From \eqref{SETBYAB} we find that $\Braket{T}=0$, $\braket{\overline{T}}=0$, and $\Braket{\Theta}=\frac{1}{8\pi G}\sin^2\phi$, confirming the identity holds.

In a $T\overline{T}$-deformed CFT, the expectation values in a stationary state obey
\be\label{mupar}
\Braket{T^a_a} = -4\pi \mu \langle T\overline{T} \rangle.
\ee
Although we are considering a holographic RG flow rather than a $T\overline{T}$-deformed boundary theory, one may nevertheless use this relation to define an effective, scale‑dependent parameter $\mu(\phi)$.


Using \eqref{mupar} together with \eqref{traceset} and \eqref{ToverT}, we obtain an effective, scale‑dependent parameter
\begin{equation}
\mu(\phi(w_c)) = \frac{8G}{\sin^2\phi(w_c)} = 8G\bigl(1+e^{-8a^2 w_c}\bigr),
\end{equation}
where \( w_c \) denotes the location of the cutoff surface. This function decreases monotonically along the RG flow: in the UV limit \( w_c \to -\infty \) (equivalently \( \phi \to 0 \)) we have \( \mu \to +\infty \), while in the IR \( w_c \to +\infty \) (\( \phi \to \pi/2 \)) we obtain the finite value \( \mu \to 8G \). Notably, \( \mu \) does not depend on the parameter \( a^2 \). A similar monotonic behavior is exhibited by the holographic \( c \)-function 
\be
c \sim 1/\cos^2\phi \propto 1/A', 
\ee
in accordance with the \( c \)-theorem.

\section{Discussion}
\label{sec:disc}
In this work we have constructed solutions with AdS, dS and Minkowski asymptotics in a three-dimensional truncated supergravity coupled to a sigma model with target space $S^2=SU(2)/U(1)$. The action includes a scalar with a periodic scalar potential. 
The solutions interpolate between AdS or dS extrema of the potential to Minkowski vacuum, which also corresponds to an extremum of the potential. 
Depending on the value of $a^2$, which is related to the curvature of the target space, the scalar potential of the holographic model gives rise to both AdS and dS solutions, which are dual to CFTs.

We provide a holographic interpretation for the exact half-supersymmetric AdS-to-Minkowski solutions flowing from a minimum to minimum of the potential. With Dirichlet boundary conditions, they are dual to deformations of the two-dimensional CFTs driven by non-zero VEV of irrelevant operators, spontaneously breaking conformal symmetry. In \cite{Kiritsis:2016kog,Gursoy:2018umf}, similar solutions to holographic models with generic types of scalar potentials were discussed as holographic exotic RG flows.
The solutions from de Sitter to Minkowski are also irrelevant deformations, but the dual theory is non-unitary. For the de Sitter case, the sign of the "VEV" factor affects the asymptotic behavior of the flows near Minkowski. We have found that it can be either asymptotically periodic or linear.

Doing a  generalization to  the finite temperature case, we have found that the only regular geometries in this case are BTZ and SdS black  holes, the other thermal solutions are singular. This seems to be in agreement with the analysis of \cite{Gursoy:2018umf}.

We have also discussed  the black string solution \cite{Deger:1999st} to our supergravity model. For which we have shown that it can be associated with a deformation triggered by a non-zero VEV of an irrelevant operator, such that the VEV is related to the mass of the black string.  An interesting point is that the black string solution is extremal ($T=0$) and, after a suitable change of radial coordinate, resembles the domain wall solution interpolating from AdS to Minkowski, while also satisfying the first-order BPS equations.

Reducing the equations of motion to an autonomous dynamical system defined on various phase spaces (on the plane, in the disk, in the cylinder and in the ball), we have clarified the behavior of the flows for certain regions.  We have also provided an analytical description  of the solution near the horizon region. We have shown that the initial conditions for the scalar field on the horizon determine the behavior of the thermal solutions.

We have explicitly computed the Brown-York stress-energy tensor on a cutoff surface at $w = w_c$ for the exact holographic RG flow and demonstrated the factorization of the  $T\overline{T}$-operator  along the flow.  Furthermore, we have introduced an effective, scale-dependent deformation parameter $\mu\big(\phi(w_c)\big)$ whose running is governed by the scalar field on the cutoff surface.

An important future direction is to extend this analysis to higher-dimensional settings, particularly through the uplift of our three-dimensional gauged supergravity solutions to string and M-theory models.  It is natural to ask whether our AdS$_3$ and dS$_3$ solutions can be realized as geometries of wrapped $M2$- or $M5$-branes, or as compactifications of AdS$_4$ or AdS$_5$ supergravity with suitable internal manifolds.
Such an uplift would clarify the microscopic interpretation of the exotic RG flows and finite-temperature solutions we have observed. In particular, the black string solution we discussed may be related to a deformed $M5$-brane or $NS5$-brane system, where the VEV of the irrelevant operator finds a natural interpretation in terms of brane moduli or worldvolume fluxes \cite{Pilch:2000fu}.

For future research, it would also be interesting to apply this dynamical systems approach to five-dimensional gauged supergravity models. A promising candidate is the theory in \cite{Bobev:2020lsk}, where the full $D=5$, $\mathcal{N}=8$ gauged supergravity is truncated to a two-scalar model. A comprehensive study of the RG flow phase portrait for this theory is clearly desirable. Another relevant model, constructed for $D=5$, $\mathcal{N}=4$ gauged supergravity, can be found in \cite{Karndumri:2022dtb}. We hope to address these models in future work.

An interesting direction of future research is to extend our analysis  to the case of  other exotic RG flows, such as limit cycles in the context of certain holographic models.  Despite the substantial amount of work on these non-perturbative effects (for example, from the perspective of gravity, see \cite{Kiritsis:2016kog,Gursoy:2018umf,Bea:2018whf}), the question of the fundamental nature and physical understanding of such non-perturbative effects remains open. To understand the nature of such effects, it is necessary to consider a large number of model examples in which exotic flows can exist.

It would also be intriguing to explore the scalar sector of 
$D=3$, $\mathcal{N}=2$ 
gauged supergravity extended with a Fayet–Iliopoulos term \cite{Abou-Zeid:2001inc}. Beyond the supersymmetric solutions already known in this context \cite{Deger:2025vni}, a particularly interesting direction is the search for gravitational instantons—vacuum solutions that satisfy the Einstein equations with vanishing energy-momentum tensor, as studied in \cite{deHaro:2006ymc,deHaro:2014xfa}. Uncovering the connection between such gravitational configurations on the bulk side and RG flows on the dual QFT side remains an open problem for future investigation.

\section*{Acknowledgments}
The authors would like to thank E. T. Musaev and N. S. Deger for helpful discussions. LA thanks Boğaziçi University and especially I. Gahramanov for kind hospitality where a part of this work was done.

\appendix

\setcounter{equation}{0} \renewcommand{\theequation}{A.\arabic{equation}}
\section{Solutions to the model}
\subsection{dS, AdS, Minkowski and SdS solutions}
\label{appendix A}
Here we consider the case $f=1$.
Combining eqs. (\ref{eom1})-(\ref{eom3}) we have 
\be\label{scalefactor}
\ddot{A}+\frac{\dot{\phi}^{2}}{a^2}=0.
\ee

Assuming that the  scalar field is almost constant leads to $\frac{\dot{\phi}^{2}}{a^2}\sim0$, so
integrating (\ref{scalefactor})  we obtain
\be
A =c_1w +c_2.
\ee

At the same time, from (\ref{eom3}) we find the following scale factor: 
\be
A =\sqrt{-2V(\phi_{*})}w+\mathrm{c}_2.
\ee
For  de Sitter we have
\be
V=\frac{2 a^4}{1+2a^2},
\ee
i.e.
\be
A=\frac{2a^2}{\sqrt{1+2a^2}}iw
\ee
with $c_2=0$.

Then the dS metric can be represented in the following form:
\be
ds^2=e^{\frac{4a^2}{\sqrt{1+2a^2}}iw}\left(-dt^{2}+dx^2\right)+dw^2,
\ee
with the constant scalar field
\be
\phi = \phi_{dS}.
\ee
One can reach the standard form of the de Sitter metric using double Wick rotation
\be
\label{dSdoublewick}
t\to -ir,\quad -dt^2=dr^2,\quad  w\to -i\tau, \quad dw^2\to-d\tau^2,
\ee
this yields the metric
\be\label{dSfsc}
ds^2 = -d\tau^2 + e^{a^{2}\sqrt{\frac{1}{1+2a^2}}\tau}\left(dr^2 + dx^2\right).
\ee
It seems that these coordinates are called "flat slicing".

Note that at the same time, de Sitter spacetime can be represented in the following form (in the static coordinates):
\be
ds^2 = -\big(1-br^2\big) dt^2+\frac{dr^2}{1-br^2} +r^2d\varphi^2,
\ee
where $b$ is a positive constant. The  cosmological horizon is located at $r=b^{-1/2}$.

The Schwarzschild-de Sitter solution is given by the following metric in the static coordinates \cite{Spradlin:2001pw}:
\be
ds^2=-\big(1-2m-r^2\big)dt^2 +\frac{dr^2}{1-2m-r^2}+r^2d\varphi^2,
\ee
where $m=4GE$. The horizon is located at $r_{h}=\sqrt{1-2m}$, and the radial coordinate changes as $0<r<\sqrt{1-2m}$.

For the Minkowski solution, we have that the potential vanishes, $V=0$. The scale factor in this case is just zero, $A=0$. So the metric is
\be
ds^2 = -dt^2 +dx^2+dw^2.
\ee

\subsection{Asymptotic behavior of flows from linearized EOMs}

In this subsection we discuss how to find solutions to eqs. (\ref{eqsA})-(\ref{eqd}) analytically, assuming that we are in the region of the extrema of the potential. In particular, we find an asymptotically periodic solution near Minkowski.

\subsubsection{Flows near $AdS$ and Minkowski extrema}

We start from the $AdS$ extremum with $\phi=0$, so the equations of motion (\ref{eqsA})-(\ref{eqd}) take the following form:
\bea\label{eqA}
\ddot{A}+2\dot{A}^2-8&=&0,\\ \label{eqA2}
\ddot{\phi}+2\dot{A}\dot{\phi}-2a^2V_{\phi}&=&0,
\eea
where we used that the linearized potential near the AdS extremum is given by
\be
V=-2,\quad V_{\phi}=V_{\phi\phi}(\phi_{\star})\cdot\phi, \quad V_{\phi\phi}=16.
\ee

For eqs. (\ref{eqA})-(\ref{eqA2}), the scale factor has a special solution given by
\be\label{partialsol}
A=2w.
\ee
The asymptotic behavior of the scale factor \eqref{partialsol} goes to infinity $A\to -\infty$, as $w
\to -\infty$, that match with the scale for the exact half-supersymmetric solution \eqref{Degsol} near AdS.

The solution to the scalar field equation (\ref{eqA2}) in this case reads
\be\label{solphi_vphi0neq}
\phi = c_1 e^{-8w}+c_2e^{4w},
\ee
where $c_1$ and $c_2$ are constants of integration. This solution is consistent with the behavior of the exact solution near AdS \eqref{Degsol} if one chooses $c_1=0$ in \eqref{solphi_vphi0neq}. 

We see that  the reconstructed solution is in agreement with the Dirichlet boundary conditions 
\be
\phi_{-}=0.
\ee

Now we turn to the Minkowski extremum $\phi =\frac{\pi}{2}$, for which the potential and its derivative are
\be
V=0,\quad V_{\phi} = 8\left(\phi-\frac{\pi}{2}\right).
\ee
The solution for the scale factor $A$ is trivial, 
\be
A =0.
\ee
The equation for the scalar field near $\phi =\frac{\pi}{2}$ reads
\be
\ddot{\phi}- 16\left(\phi-\frac{\pi}{2}\right)=0
\ee
and has the following solution:
\be\label{phiMinksol}
\phi =\frac{\pi}{2}+c_{1}e^{4w}+ c_{2}e^{-4w}.
\ee

The exact solution has the Minkowski asymptotics for $w\to +\infty$. The solution (\ref{phiMinksol}) is in agreement with this; if one sets  $c_1=0$, then we reach $\phi=\frac{\pi}{2}$. Thus, we have reproduced the asymptotic behavior of the exact solution \eqref{Degsol} correctly. Below, we will show that the solutions to the equations of motion near extrema of the potential match with those to dynamical system equations near equilibria points.

\subsubsection{Flows near $dS$ extrema}
\label{sec:dSsol}
Similarly, we can find the asymptotic behavior near $dS$ extrema. For this case, the equations of motion (\ref{eqsA})-(\ref{eqd}) take the following form:
\bea\label{eqA22}
\ddot{A}+2\dot{A}^2+4\frac{2 a^4}{1 + 2 a^2}&=&0,\\
\ddot{\phi}+2\dot{A}\dot{\phi}-2a^2V_{\phi}&=&0,
\eea
where we used that the linearized potential near the dS extremum is given by
\be
V=\frac{2 a^4}{1 + 2 a^2},\quad V_{\phi}=V_{\phi\phi}(\phi_{dS})\cdot(\phi-\phi_{dS}), \quad V_{\phi\phi}=-\frac{16 a^2( 1+ a^2)}{1 + 2 a^2}.
\ee
We have a special solution for the scale factor given by
\be\label{specsoldS}
A=\pm\frac{i\,2a^2}{\sqrt{1 + 2 a^2}}w=\mathcal{C}_{A}w.
\ee

The solution for the scalar field in this case is 
\be\label{phiphids}
\phi = c_1 e^{w \left(-\sqrt{\mathcal{C}_A^2+2a^2V_{\phi\phi}}-\mathcal{C}_A\right)}+c_2 e^{w \left(\sqrt{\mathcal{C}_A^2+2a^2V_{\phi\phi}}-\mathcal{C}_A\right)}+\phi_{dS},
\ee
where
\be
\phi_{dS}=\pm \arccos\left(\sqrt{\frac{a^2}{1+2a^2}}\right), \quad \mathcal{C}^2_{A}= -\frac{4a^4}{1+2a^2}.
\ee
One can see that the solution \eqref{phiphids} is complex in terms of the domain wall coordinates, since
\be
-\mathcal{C}_{A}\pm\sqrt{\mathcal{C}^2_{A}+2a^2V_{\phi\phi}}= \frac{2a^2i}{\sqrt{1+2a^2}}\left(\pm\sqrt{9+8 a^2}-1\right).
\ee
However, the solution for $\phi$ in terms of $A$ is real and can be represented as follows:
\be
\label{lindSsol}
\phi = c_1 e^{\left(-\sqrt{9+8a^2}-1\right)A}+c_2 e^{\left(\sqrt{9+8a^2}-1\right)A}\pm \arccos\left(\sqrt{\frac{a^2}{1+2a^2}}\right).
\ee
Below, we will also show that the solution near dS \eqref{lindSsol} can be recovered from the analysis of the dynamical system \eqref{dSdynsys}. The solution for $dS$ with $f=1$ has the complex scale factor \eqref{specsoldS}, but doing double Wick rotation, we come to the dS metric in an expanding patch (see \eqref{dSdoublewick} and \eqref{dSfsc}). If we set $f=-1$ from the start, we obtain the real $dS$ metric in the form \eqref{dSfcoord} accurate to just the redefinition of coordinates corresponding also to the dS expanding patch.

\setcounter{equation}{0} \renewcommand{\theequation}{B.\arabic{equation}}

\section{Lyapunov analysis}
\label{appendix B}
Implement the Lyapunov analysis on the dynamical system. Define the functions
\be
f=X,\quad
g=\left(\frac{X^2}{a^2}-2\right)\left(X+\frac{a^2}{2}\frac{V'}{V}\right).
\ee
and its derivatives matrix
\bea
\mathcal{M}= \left( \begin{array}{cc}
\frac{\partial f}{\partial \phi} & \frac{\partial f}{\partial X}  \\
\frac{\partial g}{\partial \phi} & \frac{\partial g}{\partial X} \end{array} \right)\Big|_{X=X_c,\phi=\phi_c},
\eea
Demonstrate the analysis explicitly for the dS point. This point has the following coordinates:
\be X_c=0,\quad
\phi_{c} = \phi_{dS} = \pm\arccos\left(\sqrt{\frac{a^2}{1+2a^2}}\right
).\ee
The matrix elements are as follows:

\bea
\mathcal{M}_{11}&=&0;\quad \mathcal{M}_{12}=1;\\
\mathcal{M}_{21}&=&-\frac{\sec^2\phi\big(X^2-2\big)\big(13+3\cos4\phi\big)}{\big(1-3\cos2\phi\big)^2}\Big|_{X=X_c,\phi=\phi_c}=8 (1 + a^2);\nonumber\\
\mathcal{M}_{22}&=&\frac{3 X^2}{a^2}-\frac{2 X \tan\phi \big(\left(2 a^2+1\right) \cos2\phi+1\big)}{\big(2 a^2+1\big) \cos ^2\phi-2
   a^2}-2\Big|_{X=X_c,\phi=\phi_c}=-2;\nonumber
\eea
\be
\det\mathcal{M}=-16.
\ee
The characteristic equation
\be
\lambda^2 -\lambda\big(\mathcal{M}_{11}+\mathcal{M}_{22}\big)+\mathcal{M}_{11}\mathcal{M}_{22}-\mathcal{M}_{12}\mathcal{M}_{21}
=0\ee
has the following solutions:
\be
\lambda_{1,2}=-1\pm\sqrt{9+8 a^2}.
\ee
Hence, we conclude that the de Sitter fixed point is a saddle.

\begin{figure}[ht]
    \centering
\includegraphics[height=10cm]{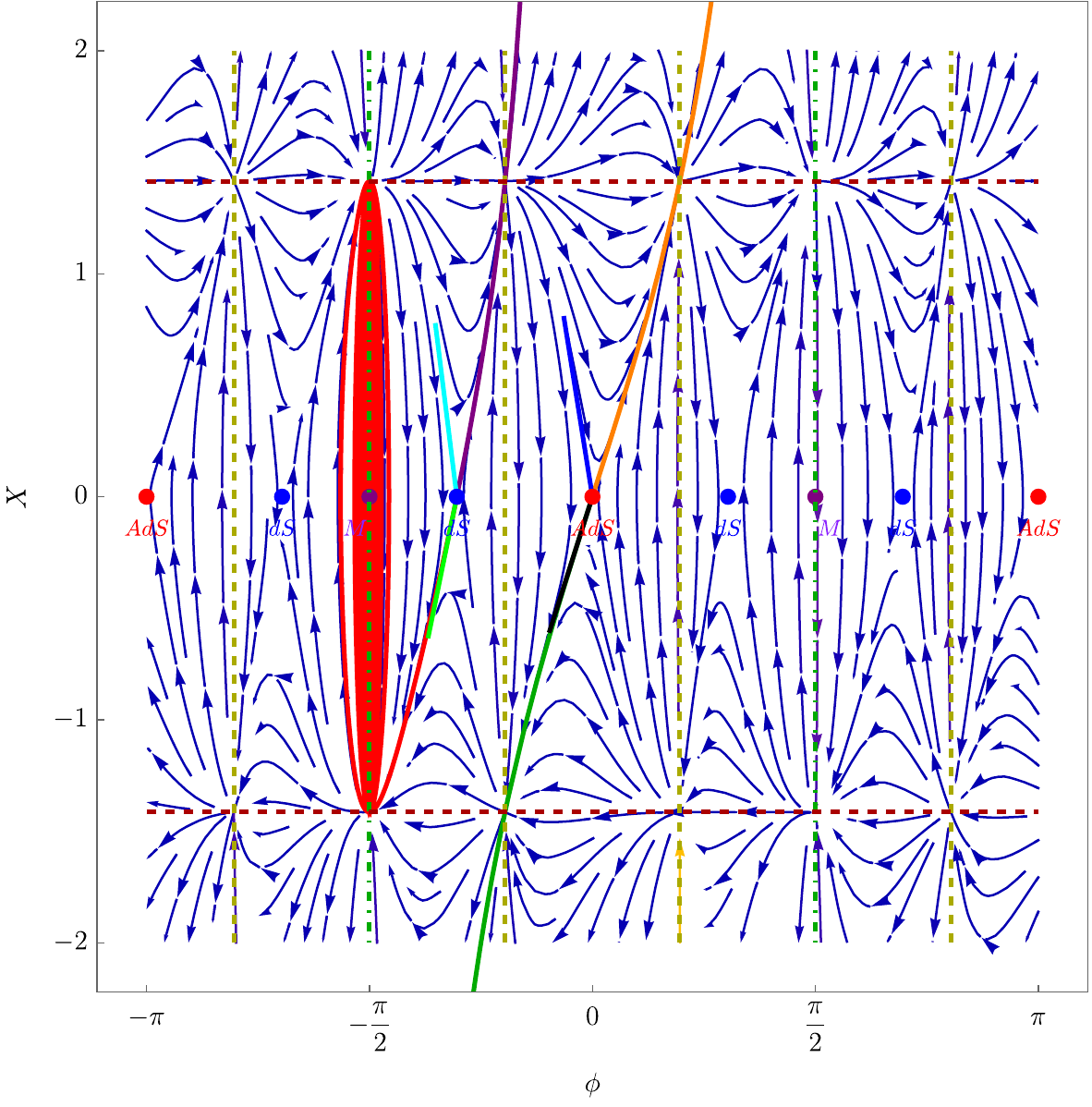}
    \caption{Phase flows from AdS and dS to Minkowski with the depicted eigenvectors directions. Blue and black straight lines correspond to $AdS$ eigenvectors, and cyan and green straight lines correspond to $dS$'s.}
\label{fig:flowseigen}
\end{figure}

The eigenvectors are defined by the following equation:
\be
\mathcal{M}^{i}_{\, j}u^{j}_{1,2}=\lambda_{1,2}u_{1,2}^{i},
\ee
from where we can find them
\be
u_{1} =\left[\begin{array}{c}
1\\
\lambda_{1}\end{array}\right],\quad u_{2} =\left[\begin{array}{c}
1\\
\lambda_{2}\end{array}\right].
\ee
We know that eigenvalues of the Lyapunov exponent correspond to the operators scaling dimensions with the opposite sign: 
\be
\lambda_{1,2}=-\Delta_{1,2} \, .
\ee
Hence, we can obtain the solution near dS fixed points, which is given by
\bea\label{dSdynsys}
\left[\begin{array}{c}
\phi\\
X
\end{array}\right]=\left[\begin{array}{c}
\pm\arccos\left(\sqrt{\frac{a^2}{1+2a^2}}\right)\\
0
\end{array}\right]+k_{1}e^{-\Delta_{1}A}u_{1}+k_{2}e^{-\Delta_{2}A}u_{2}.
\eea
In the same manner, the general solution for the AdS point at $X_c=0$ and $\phi_c = \phi_{AdS} = 0$ is given by 
\bea
\left[\begin{array}{c}
\phi\\
X
\end{array}\right]=\left[\begin{array}{c}
0\\
0
\end{array}\right]+k_{1}e^{-\Delta_{1}A}u_{1}+k_{2}e^{-\Delta_{2}A}u_{2}
\eea
with the eigenvalues $\lambda_{1,2}= - \Delta_{1,2} = -1\pm (1+2a^2)$ and the eigenvectors
\be
u_{1} =\left[\begin{array}{c}
1\\
\lambda_{1}\end{array}\right],\quad u_{2} =\left[\begin{array}{c}
1\\
\lambda_{2}\end{array}\right].
\ee
We see that the $AdS$ point is also a saddle. In Fig.~\ref{fig:flowseigen}, we draw the directions of the eigenvalues corresponding to the central $AdS$ point and to its closest left $dS$ point.

\newpage
\bibliography{bib.bib}
\bibliographystyle{utphys.bst}

\end{document}